\documentclass{article}
\usepackage{graphicx} % Required for inserting images

\usepackage{comment}  % put in preamble

\usepackage{PRIMEarxiv}

\usepackage[utf8]{inputenc} % allow utf-8 input
\usepackage[T1]{fontenc}    % use 8-bit T1 fonts
\usepackage{hyperref}
\usepackage{url}            % simple URL typesetting
\usepackage{booktabs}       % professional-quality tables
\usepackage{bm}

\usepackage{soul}
\usepackage{comment}
\usepackage{stmaryrd}
\usepackage{amsfonts}       % blackboard math symbols
\usepackage{amsmath}
\usepackage{amssymb}
\usepackage{authblk}
\usepackage{nicefrac}       % compact symbols for 1/2, etc.
\usepackage{microtype}
\usepackage{cleveref}
\usepackage{xcolor}
\usepackage{algpseudocode}
\usepackage{algorithm}
\usepackage{color,soul}
\usepackage{dsfont}
\usepackage{breqn}
\usepackage{caption}
\usepackage{lipsum}
\usepackage{fancyhdr}       % header
\usepackage{graphicx}       % graphics
\graphicspath{{media/}}     % organize your images and other figures under media/ folder
\usepackage{enumitem}
\usepackage{empheq}
\makeatletter
\newcommand{\printfnsymbol}[1]{%
  \textsuperscript{\@fnsymbol{#1}}%
}

\makeatother
\title{VASTO: Simultaneous reconstruction of vascular geometry and blood flow from tomographic projections via differentiable topology optimization}

\author[1]{Pramod Thombre}
\author[1]{Rahul Kumar Padhy}
\author[2]{Roshan M. D'Souza}
\author[1,*]{Krishnan Suresh}

\affil[1]{Department of Mechanical Engineering, University of Wisconsin-Madison, Madison, WI, USA 
\authorcr
   \{\tt pthombre, rkpadhy, ksuresh\}@wisc.edu}

\affil[2]{Department of Mechanical Engineering, University of Wisconsin-Milwaukee, Milwaukee, WI, USA 
\authorcr
   \tt dsouza@uwm.edu}
   
\affil[*]{Corresponding author}
\date{December 2025}

\begin{document}
\maketitle

\begin{abstract}

Computed Tomography Angiography (CTA) is widely used to reconstruct vascular geometry from projection measurements, with conventional approaches such as Filtered Back-Projection (FBP) and Iterative Reconstruction (IR) forming the clinical standard. Despite recent advances, these methods remain susceptible to sparse and noisy measurement, making accurate recovery of critical anatomical features challenging. This consequently leads to poor estimation of blood flow. In this work, we present a \textit{fluid-physics-constrained} reconstruction framework that leverages topology optimization (TO) to jointly recover vascular geometry and blood flow directly from time-resolved CTA sinograms.  The formulation couples a steady incompressible flow model with a transient advection-diffusion contrast transport model, mapped to sinogram space through a differentiable projection operator. The velocity field is recovered as part of the reconstruction, and thus hemodynamic information can be obtained without requiring a separate computational fluid dynamics pipeline. The proposed method is demonstrated on synthetic phantoms under varying sparsity and noise levels, and on representative projection data.

\end{abstract}

    \begin{figure}[htbp]
        \centering
            \includegraphics[scale=0.38,trim={0 0 0 0},clip]{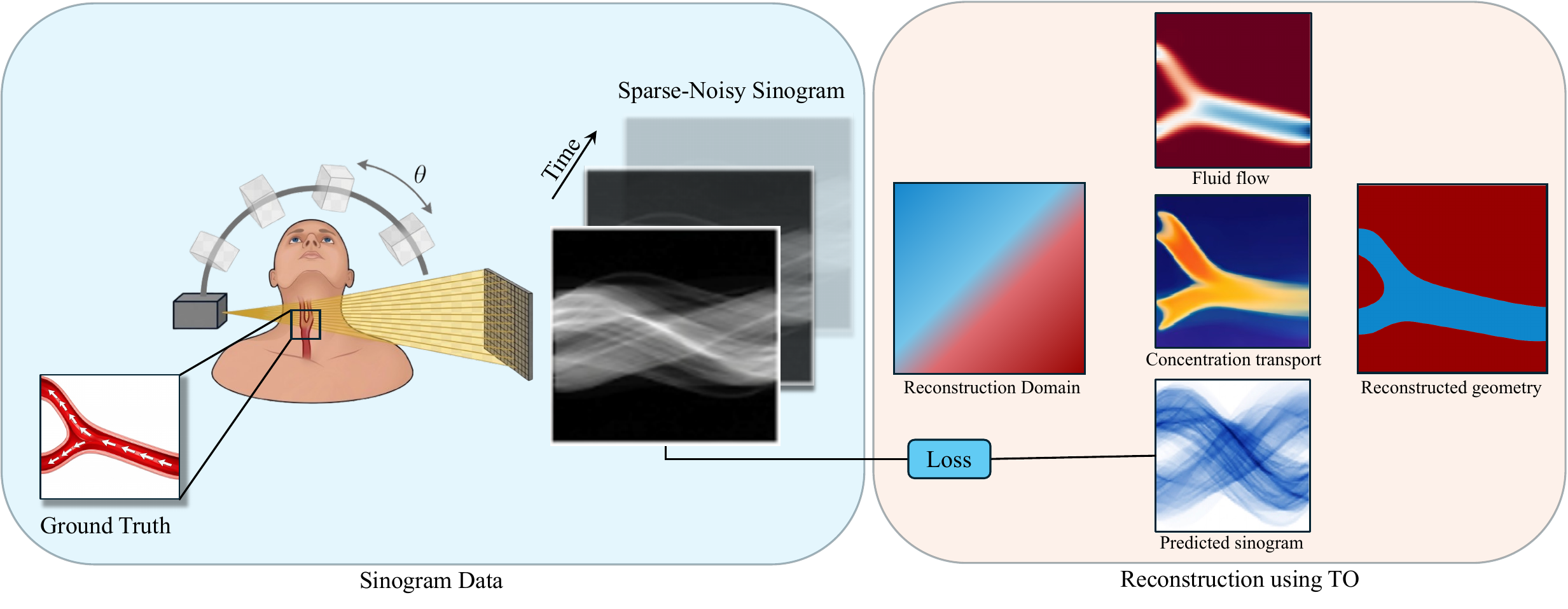}
            \caption*{Graphical abstract of the proposed framework: the predicted sinograms generated from flow and transient transport simulations are matched with sparse CT scan data (sinogram) to reconstruct the vascular geometry and the blood flow via topology optimization.}
    \end{figure}

\section{Introduction}        

Cardiovascular disease (CVD) is a leading cause of mortality and disability worldwide \cite{global2025global, hong2025comprehensive}. Clinical assessment of CVD relies critically on hemodynamics, the analysis of blood flow within vascular structures \cite{secomb2016hemodynamics, keshavarz2020diagnostic, pinsky2022effective}. In particular, hemodynamic descriptors such as Fractional Flow Reserve (FFR) and Wall Shear Stress (WSS) are critical for guiding patient-specific treatment decisions \cite{gao2025wall, stokes2023impact, bakker2023preclinical}. Among established clinical workflows, the non-invasive derivation of these descriptors relies on vascular geometries reconstructed from medical imaging\cite{stokes2023impact}. Accurate reconstruction of patient-specific vascular geometry is essential for robust hemodynamic analysis \cite{bovsnjak2025geometric, nolte2022inverse, gao2025wall}.

\begin{figure}[htbp]
        \begin{centering}
            \includegraphics[scale=0.45,trim={0 0 0 0},clip]{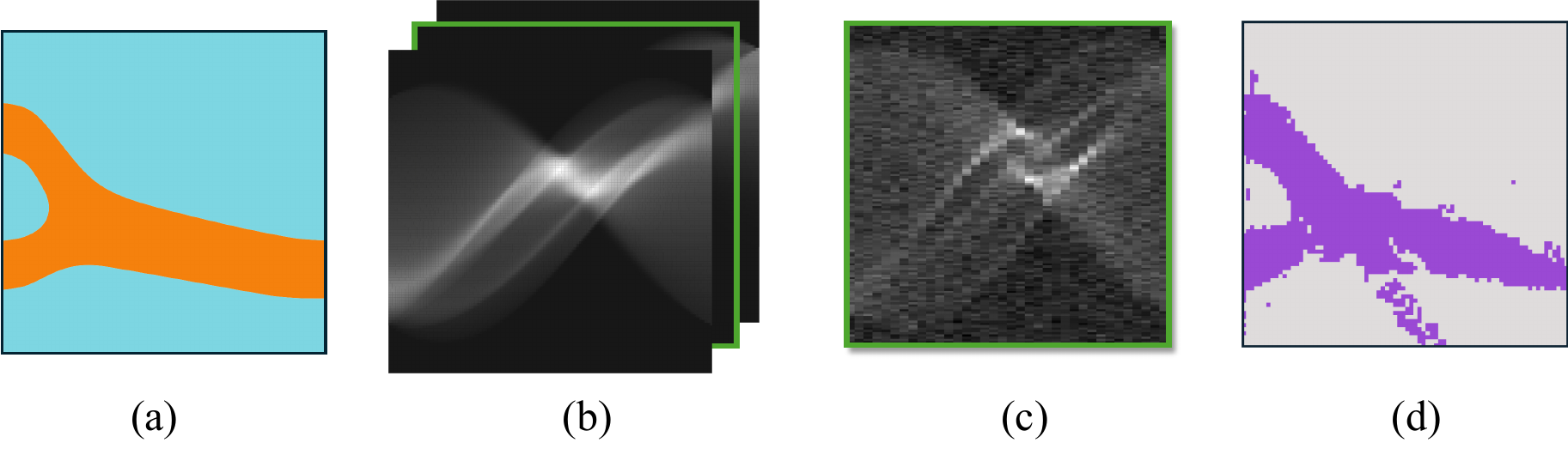}
            \caption{Conventional CT pipeline on a 2D bifurcating artery phantom with sparse-noisy acquisition. (a) Phantom 2D geometry of a bifurcated artery used as a ground truth. (b) Sinogram data. (c) Sparse and noisy sinogram data.  (d) Reconstructed lumen (geometry) using FBP.}
            \label{fig:introduction_fbp}
        \end{centering}
    \end{figure}

Contrast-enhanced computed tomography (CE-CT) is a standard imaging modality for reconstructing vascular geometries \cite{sixou2020contrast}. Conventional CE-CT pipelines reconstruct CT images from projection data using filtered back-projection (FBP) \cite{van2014scikit} or iterative methods, followed by lumen segmentation to extract geometry \cite{herman2009fundamentals, yao2025physics}.  Despite advances in deep-learning architectures \cite{yao2025physics}, these reconstruction methods remain susceptible to sparse and noisy measurements characteristic of time-resolved acquisitions \cite{yao2025physics, shusong2024deep, guo2025computed}. Specifically, these limitations arise from radiation dose constraints and physiological motion artifacts inherent in arterial imaging \cite{nolte2022inverse, fahrni2023investigating, guo2025computed}. These limitations result in undersampling artifacts, degrading recovered geometry as shown in \cref{fig:introduction_fbp}.

To improve the robustness of flow reconstruction in presence of data limitations, recent frameworks incorporate governing physics into hemodynamic parameter estimation (e.g., velocity, flow rate, pressure, and boundary conditions) \cite{sixou2020contrast, bakker2021image, guo2025computed, raissi2020hidden}. These methods typically couple the Navier–Stokes equations and the advection–diffusion equation \cite{sixou2020contrast, bakker2021image, guo2025computed, raissi2020hidden, bakker2023preclinical} to model blood flow and contrast transport dynamics. However, a key limitation is that these methods often assume a fixed, pre-defined geometry and focus solely on estimating flow parameters or boundary conditions \cite{nolte2022inverse, sixou2020contrast, bakker2021image, guo2025computed}. Enforcing flow physics on a pre-defined geometry prior yields a valid but unrepresentative solution, failing to capture the critical hemodynamic parameters. Rather than prescribing the vascular geometry a priori, \cite{kontogiannis2022joint} pose an inverse Navier–Stokes problem to jointly reconstruct noisy velocity images and infer the flow boundary. However, their framework assumes that the input data are measured velocity images, and therefore does not address inverse problems where the geometry must be recovered directly from projection measurements.

In this paper, we introduce a physics-constrained reconstruction framework based on topology optimization (TO) \cite{bendsoe1988generating}. Adopting density-based TO \cite{borrvall2003topology, alexandersen2023detailed, padhy2026toflux}, we treat the vascular geometry as an unknown rather than a pre-segmented input. Our objective is to reconstruct a two-dimensional vascular lumen geometry directly from time-resolved CE-CT sinograms by solving a physics-constrained inverse problem. Specifically, the geometry enters the incompressible Navier--Stokes equations through a Brinkman penalization \cite{alexandersen2023detailed, padhy2026toflux}, and the resulting velocity field drives a transient advection--diffusion model of the contrast agent. A differentiable projection operator \cite{guo2025computed} maps the concentration fields of the contrast agent to sinogram space, so that the geometry is optimized until its flow and contrast transport reproduce the measured time-resolved sinograms. This unified formulation eliminates the dependence on a fixed a priori domain, ensuring a robust reconstruction even under the constraints of sparse and noisy data. 

The remainder of the paper is organized as follows. Related work is reviewed in \cref{sec:related_work}. The formulation, discretization and optimization are presented in \cref{sec:proposed_method}, and the numerical experiments in \cref{sec:numerical_experiments}. The results are discussed in \cref{sec:discussion}. Future work is discussed in \cref{sec:futurework}, and conclusions are drawn in \cref{sec:conclusion}.

\section{Related Work}
\label{sec:related_work}

This section reviews the work on image-based flow estimation and computational inverse design. The survey is structured around the central contribution of this paper: the reconstruction of vascular topology from tomographic projection data. We organize the review into two primary sections. The first section reviews existing frameworks in inverse hemodynamics, highlighting their reliance on fixed geometric priors and the resulting limitations. The second section examines the evolution of density-based multiphysics fluid topology optimization (TO) and its emerging biomedical applications, outlining its unexplored potential to address the geometric bottlenecks in current blood flow reconstruction pipelines. 

    \subsection{Inverse Hemodynamics}
    In recent years, the formulation of inverse problems has emerged as a powerful tool to extract unobserved hemodynamic quantities such as pressure gradient, wall shear stress (WSS), and fractional flow reserve (FFR) directly from medical imaging data \cite{nolte2022inverse, gao2025wall, stokes2023impact}. To estimate these functional parameters, existing frameworks typically couple the Navier--Stokes equations with an advection-diffusion transport model to track the dynamics of a contrast agent \cite{bakker2021image, bakker2023preclinical}. For instance, \cite{sixou2020contrast} and \cite{huang2023pod} utilized contrast-enhanced computed tomography (CE-CT) projections to reconstruct the velocity field and convection-diffusion physics using adjoint methods \cite{sixou2020contrast, huang2023pod, shusong2024deep}. Similarly, \cite{bakker2021image, bakker2023preclinical} developed a flow estimation method based on semi-analytical solutions to the advection-diffusion equation to quantify coronary blood flow from CCTA images. More recently, machine learning has revolutionized inverse hemodynamics. Building upon foundational frameworks such as hidden fluid mechanics (HFM) \cite{raissi2020hidden}, physics-informed neural networks (PINNs) have been deployed by \cite{guo2025computed} (SinoFlow) to infer blood flow velocity directly from CT sinograms by minimizing residuals of the governing physical equations \cite{guo2025computed}. 
    
    While these methods successfully recover complex hemodynamics, they share a critical limitation: they fundamentally rely on an \textit{a priori}, fixed geometric representation of the blood vessel \cite{nolte2022inverse, bovsnjak2025geometric, kontogiannis2022joint}. In conventional pipelines, this geometry is extracted sequentially, first by reconstructing a CT volume using FBP or deep learning, and subsequently segmenting the vessels using techniques ranging from classical thresholding to advanced 3D U-Net architectures \cite{sadid2022segmenting, wang2023automatic}. However, under ultra-sparse acquisitions \cite{yao2025physics} or in the presence of motion artifacts from fast CT gantry rotations \cite{fahrni2023investigating}, this disjointed segmentation process introduces severe geometric uncertainties, such as missing branches or artificial stenoses. As recently demonstrated by \cite{bovsnjak2025geometric}, hemodynamic biomarkers such as WSS and pressure drops are highly sensitive to even minor local boundary perturbations \cite{bovsnjak2025geometric, stokes2023impact}. Consequently, enforcing flow physics on an erroneous fixed geometry yields a valid but unrepresentative flow field. Recognizing this limitation, the authors of \cite{kontogiannis2022joint} formulated a generalized inverse Navier-Stokes problem for MRI velocity data, demonstrating that treating the vessel geometry as an unknown optimization variable is necessary to simultaneously reconstruct flow and correct segmentation errors. Motivated by this work, the present study seeks to recover geometry from sparse CE-CT projections by formulating the geometry recovery as a density-based fluid topology optimization problem, the foundational developments of which are reviewed next.
    
    \subsection{Multiphysics Topology Optimization}
    
    Treating geometry as an unknown in inverse problems has a rich history, particularly through the use of level-set methods to reconstruct obstacles and evolving domain boundaries \cite{santosa1996level, burger2005survey}. In the context of medical imaging, this strategy has driven the development of simultaneous reconstruction and segmentation (SRS) frameworks \cite{romanov2016simultaneous, yoon2010simultaneous}. By coupling the reconstruction process with geometric priors, SRS methods avoid the error propagation inherent in sequential pipelines, allowing for the recovery of sharp, topologically consistent boundaries directly from limited or noisy tomographic data \cite{romanov2016simultaneous, yoon2010simultaneous}. While level-set and discrete tomography approaches have demonstrated the power of joint shape-and-state recovery \cite{batenburg2011dart, romanov2016simultaneous, yoon2010simultaneous}, their application to multi-physics fluid domains often incurs prohibitive computational costs. Traditional boundary-tracking methods require higher-order numerical techniques to handle fluid interfaces, necessitating frequent reinitialization algorithms and complex velocity field extensions that severely restrict computational efficiency \cite{deng2013topology, burger2005survey, li2023mini, aghasi2011parametric}.

     To overcome these limitations, density-based topology optimization (TO) provides a flexible computational alternative. Originally developed for structural mechanics, density-based TO has been widely extended to Stokes and Navier--Stokes fluid flows \cite{borrvall2003topology, alexandersen2023detailed}. In fluid TO, the design domain is described by a continuous pseudo-density field, and the solid regions (e.g., vascular walls) enter the governing equations through an artificial distributed resistance, the Brinkman penalty term \cite{alexandersen2023detailed, padhy2026toflux}. \cite{gersborg2005topology} interpreted this term as a Brinkman-type model of flow through a porous medium and extended the approach to the Navier--Stokes equations at low to moderate Reynolds numbers and \cite{guest2006topology} combined the Stokes equations in the fluid with Darcy's law governing flow through solid material. The approach has since been extended to flows driven by body forces \cite{deng2013topologyunsteady} and to flows on curved surfaces \cite{deng2022topology}. Level-set methods have also been used for fluid TO \cite{deng2013topology}, including capillary two-phase flows \cite{deng2017topology}; they represent the fluid--solid boundary implicitly as the zero contour of a level-set function, at the cost of solving an additional evolution equation for that function. Because density-based formulations need no boundary tracking or re-meshing, density-based fluid TO is increasingly applied to biomedical and microfluidic systems, including lab-on-a-chip devices and drug delivery mechanisms \cite{borrvall2003topology, alexandersen2023detailed, padhy2024fluto}.
    
    Recently, fluid TO  has been extended to multiphysics regime. Modern frameworks couple fluid dynamics with secondary physical interactions, such as conjugate heat transfer and transient transport phenomena \cite{subramaniam2019topology}. A major catalyst for this expansion is the adoption of differentiable programming frameworks. By leveraging automatic differentiation (AD), modern fluid TO bypasses the laborious manual derivation of adjoint sensitivities for coupled systems \cite{padhy2026toflux}. This end-to-end differentiability enables robust gradient-based optimization for highly non-linear fluid phenomena, including non-Newtonian blood rheology \cite{padhy2026toflux}.
    
    However, despite these computational advancements, the application of density-based fluid TO has largely focused on the \textit{design} of engineering components \cite{padhy2024fluto, li2023mini}. While TO provides the exact mathematical mechanisms needed to continuously design complex, multiphysics domains without fixed boundaries, its potential for \textit{inverse reconstruction}, particularly vascular lumen recovery from sparse and noisy CT sinograms, remains largely unexplored in current comprehensive literature surveys of inverse hemodynamics \cite{nolte2022inverse}.

\section{Proposed Method}
\label{sec:proposed_method}
    This section formulates the physics-constrained inverse problem for reconstructing vascular lumen topology from sparse, noisy time-resolved CE-CT sinograms. First, we provide an overview of the proposed method, followed by detailed technical discussions.
    
    \subsection{Overview}
    \label{subsec:overview_proposed_method}
     
\begin{figure}[htbp]
        \begin{centering}
            \includegraphics[scale=0.65,trim={0 0 0 0},clip]{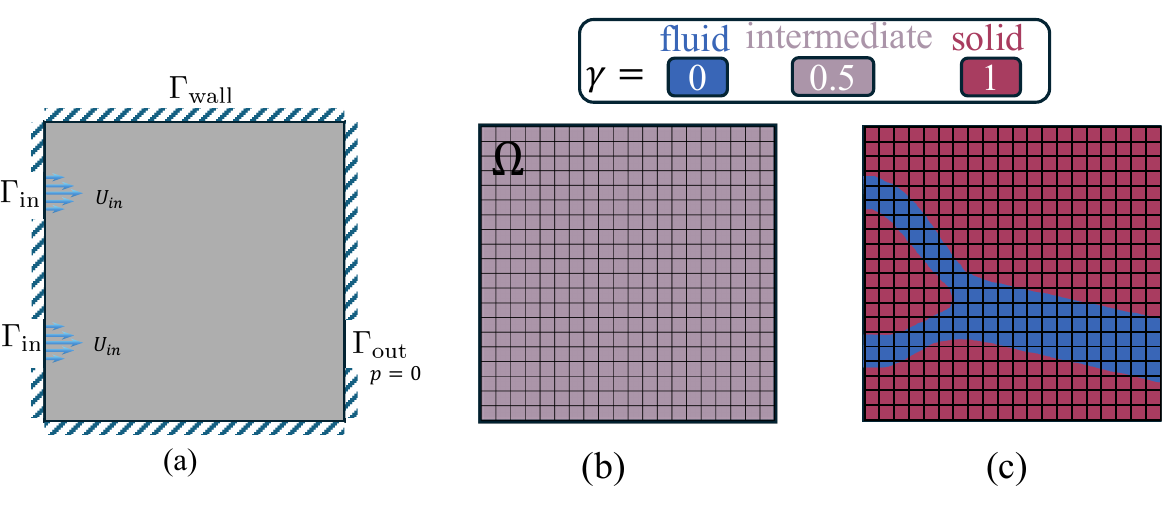}
            \caption{Design-domain representation used in the proposed topology reconstruction framework. 
            (a) Computational domain and boundary conditions. 
            (b) Initial uniform pseudo-density field, $\gamma_e=0.5$ defined on the structured mesh. 
            (c) Representative optimized pseudo density field, where $\gamma=0$ denotes fluid and $\gamma=1$ denotes solid/tissue.}
            \label{fig:proposed_method_domain}
        \end{centering}
     \end{figure}

     Consider the computational design domain in \cref{fig:proposed_method_domain}(a) where the inlet(s) and outlet(s) are prescribed but the fluid path is not known. We employ the density-based TO method to compute this fluid path \cite{borrvall2003topology}. Specifically, the domain is discretized into mesh elements, and each mesh element is assigned a pseudo-density variable $\gamma_e$, where $\gamma=0$ is interpreted as fluid, and  $\gamma=1$ is interpreted as solid. The variable is initialized to $\gamma_e=0.5$ everywhere as in \cref{fig:proposed_method_domain}(b). The objective is to determine the optimal distribution of $\boldsymbol{\gamma}$ that best explains the measured sinogram data; a plausible solution is illustrated in \cref{fig:proposed_method_domain}(c).
     
\begin{figure}[htbp]
        \begin{centering}
            \includegraphics[scale=0.9,trim={0 0 0 0},clip]{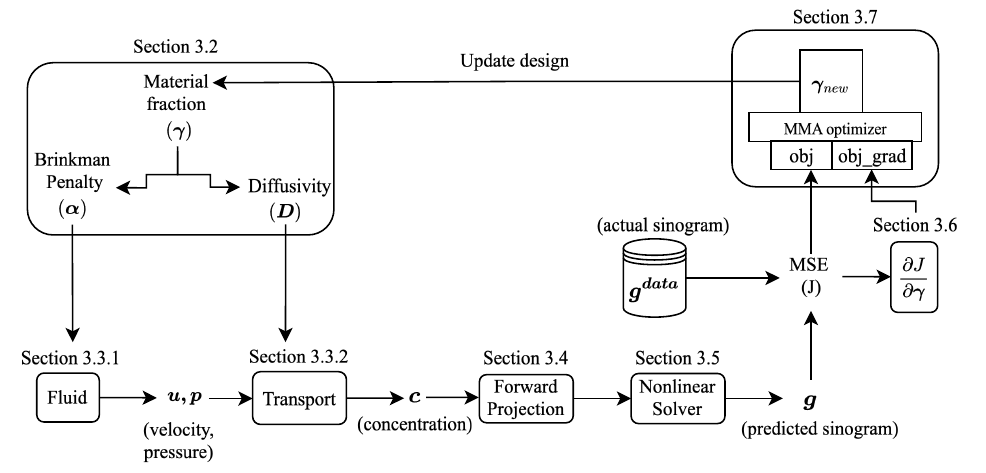}
            \caption{Optimization loop of the proposed framework.}
            \label{fig:optimization_flowchart}
        \end{centering}
    \end{figure}

    The proposed framework for determining $\boldsymbol{\gamma}$ is illustrated in \cref{fig:optimization_flowchart}. Given an initial distribution $\boldsymbol{\gamma}$, we compute design-dependent material properties, namely the Brinkman penalty $\alpha$, and the mass diffusivity $D$ as explained in \cref{subsec:design_representation}. The Brinkman penalty $\alpha$ is small in fluid regions and large in solid regions, where it suppresses the fluid velocity. Similarly, the design-dependent diffusivity $D$ limits the contrast transport in the solid regions.
    
    Using the computed Brinkman penalty $\alpha$, the steady state Navier-Stokes equations (\cref{subsec:fluid_pde}) are solved to obtain the fluid velocity field. This velocity field, together with the computed mass diffusivity $D$ is then used in the transient advection-diffusion equation (\cref{subsec:transport_pde}). This produces a time-resolved concentration field $\boldsymbol{c}$.  Next, using the forward projection operator (\cref{subsec:fwd_projection}), the simulated concentration field is projected into the measurement space to obtain the predicted sinograms $\boldsymbol{g}$. 

   Then, in \cref{subsec:optimization}, we compute the mean-squared-error (MSE) between the predicted sinograms $\boldsymbol{g}$ and the acquired data $\boldsymbol{g}^{\mathrm{data}}$. A large error indicates that the current geometry produces flow and contrast transport that are inconsistent with the acquired sinograms. We then compute the gradient of the objective with respect to $\boldsymbol{\gamma}$. Using the resulting sensitivities, the optimizer updates $\boldsymbol{\gamma}$ and repeats the loop as described in \cref{subsec:optimization}. Over successive iterations, the method searches for a fluid-solid distribution whose coupled flow, contrast transport and predicted sinograms are consistent with the acquired sinograms. Thus, the reconstructed geometry is inferred as the topology that best explains the acquired projection data through the governing physics.

    \subsection{Design Representation and Material Interpolation}
    \label{subsec:design_representation}

        The pseudo-density $\gamma_e$ determines material properties in each element. Specifically, the Brinkman penalty coefficient $\alpha$ is mapped from $\gamma$ using the Rational Approximation of Material Properties (RAMP) interpolation \cite{alexandersen2023detailed}:
        
        \begin{equation}
            \alpha(\gamma) = \alpha_{\min} + (\alpha_{\max} - \alpha_{\min}) \frac{\gamma}{1 + q_\alpha - q_\alpha \gamma},
            \label{eq:ramp_brinkman}
        \end{equation}
        
        where $\alpha_{\max}$ is a large inverse permeability that suppresses fluid flow in solid regions, $\alpha_{\min}$ is a small value retained in fluid regions to recover the original Navier--Stokes equations, and $q_{\alpha}$ controls the convexity of the interpolation profile. Similarly, the mass diffusivity $D$ in \cref{eq:transient_transport_residual} is mapped from $\gamma$ by adapting the RAMP function \cite{padhy2026tomatoes, padhy2026toflux}:
        
        \begin{equation}
            D(\gamma) = D_{\min} + (D_{\max} - D_{\min}) \frac{1-\gamma}{1 + q_D \gamma},
            \label{eq:ramp_diffusivity}
        \end{equation}
        
        where $D_{\max}$ is the physical diffusivity in the fluid ($\gamma=0$), $D_{\min}$ is a small lower bound imposed in the solid ($\gamma=1$) to avoid numerical instabilities, and $q_D$ controls the curvature of the interpolation. This ensures that the contrast agent is confined to the fluid region and prevented from artificially diffusing into the surrounding tissue.
        
        While intermediate values of $\gamma$ are permitted during optimization to facilitate gradient-based updates, a binary design is required at convergence. To this end, the pseudo-densities are first regularized using a standard density filter to ensure mesh-independence and preclude numerical artifacts such as checkerboarding \cite{sigmund1998numerical}. A subsequent smoothed Heaviside projection \cite{wu2017infill} is then applied to the filtered field, which drives the design toward a binary fluid-solid topology.
        
    \subsection{Governing Equations}
    \label{subsec:governing_equations}
    
    This section presents the governing equations of the forward model used for physics-constrained reconstruction of vascular topology.  
    
    \subsubsection{Fluid Flow}
    \label{subsec:fluid_pde}
        Specifically, blood is modeled as an incompressible, Newtonian fluid under steady-state conditions. To facilitate topology optimization, an artificial body force, the Brinkman penalty $(\alpha \mathbf{u})$ \cite{alexandersen2023detailed, padhy2026toflux}, is introduced into the Navier--Stokes equations. This term acts as a momentum sink, representing a distributed resistance that suppresses fluid velocities in solid regions. The governing equations can be expressed as:

           \begin{subequations}\label{eq:navier_stokes_system}
            \begin{empheq}
                [left={\mathcal{R}_{\mathrm{NS}}(\mathbf{u},p)=\empheqlbrace}]{align}
                \rho (\mathbf{u} \cdot \nabla) \mathbf{u}
                - \nabla \cdot \bigl(\mu(\nabla \mathbf{u} + \nabla \mathbf{u}^{T})\bigr)
                + \nabla p
                + \alpha\mathbf{u} &= 0,
                \label{eq:navier_stokes_a}\\
                \nabla \cdot \mathbf{u} &= 0,
                \label{eq:navier_stokes_b}
            \end{empheq}
        \end{subequations}

        Here, $\mathbf{u}$ is the fluid velocity vector, $p$ is the pressure, $\rho$ is the fluid mass density, $\mu$ is the dynamic viscosity, and $\alpha$ is the Brinkman penalty coefficient. A parabolic inlet velocity profile ($\bar{\mathbf{u}}$) is prescribed on the inflow boundary ($\Gamma_{\mathrm{in}}$), no-slip conditions are imposed on the remaining outer boundary ($\Gamma_{\mathrm{wall}}$), and a zero-pressure condition is applied at the outlet ($\Gamma_{\mathrm{out}}$) (\cref{fig:proposed_method_domain}(a)).

    \subsubsection{Transient Advection-Diffusion}
    \label{subsec:transport_pde}
    
    Having obtained the steady-state velocity field $\mathbf{u}$ from \cref{eq:navier_stokes_system}, we solve the transient advection-diffusion equation to simulate the transport of the injected contrast agent. The contrast agent is advected by the blood flow while simultaneously undergoing diffusion governed by the mass diffusivity $D$. The governing equation can be expressed as:
        
                \begin{equation}
                    \mathcal{R}_{\mathrm{T}}( c, \mathbf{u})
                    =
                    \frac{\partial c}{\partial t}
                    + \mathbf{u} \cdot \nabla c
                    - \nabla \cdot \left( {D(\boldsymbol{\gamma})} \nabla c \right)
                    = 0,
                    \qquad \text{in } \Omega \times (0,T],
                    \label{eq:transient_transport_residual}
                \end{equation}
        
        We assume that a time-varying Dirichlet concentration profile $c = c_{\mathrm{in}}(t)$ is prescribed at the inflow boundary ($\Gamma_{\mathrm{in}}$), representing the injection of the contrast agent. A zero diffusive-flux condition is applied on the remaining boundary; contrast leaves by advection through $\Gamma_{\mathrm{out}}$.
        
    \subsection{Forward Projection}
    \label{subsec:fwd_projection}
            
        Having computed the transient concentration field $c(x,y,t)$, the final step of the forward model maps this to the CE-CT measurement space. Specifically, we construct a time-resolved sinogram by forward-projecting the concentration field under a parallel-beam CT geometry. In this setting, each projection ray is indexed by its view angle $\theta \in [0,\pi)$ and detector coordinate $s$, with:
        \[
        \ell(\theta,s)=\{(x,y)\in\mathbb{R}^2 : x\cos\theta + y\sin\theta = s\}.
        \]
        
        Under the Beer--Lambert law \cite{goldman2007principles}, the transmitted X-ray intensity along the ray $\ell(\theta,s)$ at time $t$ is:
        \begin{equation}
            I(t,\theta,s)
            =
            I_0 \exp\!\left(-\int_{\ell(\theta,s)} \lambda(x,y,t)\,dl\right),
            \label{eq:exponential_attenuation}
        \end{equation}
        where $I_0$ is the incident intensity and $\lambda(x,y,t)$ is the linear attenuation coefficient \cite{bakker2021image}. To relate the measurements to the transported contrast agent, we assume that the attenuation varies linearly with the local concentration,
        \begin{equation}
            \lambda(x,y,t)=\lambda_{\mathrm{bg}}+\kappa\,c(x,y,t),
            \label{eq:mu_conc_link}
        \end{equation}
        where $\lambda_{\mathrm{bg}}$ denotes the static background attenuation and $\kappa$ is a proportionality constant.
        
        To isolate the dynamic contrast signal, tomographic pipelines typically employ a pre-contrast mask scan, acquired before contrast injection, given by $I_{\mathrm{mask}}(\theta,s) = I_0 \exp(-\int \lambda_{\mathrm{bg}}\,dl)$ \cite{huang2022simultaneous}. Substituting \cref{eq:mu_conc_link} into \cref{eq:exponential_attenuation}, and taking the negative logarithm of the ratio between the contrast-enhanced intensity $I(t,\theta,s)$ and the static mask intensity $I_{\mathrm{mask}}(\theta,s)$, cancels the background tissue contribution \cite{guo2025computed, sixou2020contrast}. By further normalizing this log-subtracted data by $\kappa$, we obtain a linear projection model that depends strictly on the transient fluid concentration \cite{goldman2007principles, sixou2020contrast}:
        \begin{equation}
            g(t,\theta,s)
            =
            -\frac{1}{\kappa}\ln\!\left(\frac{I(t,\theta,s)}{I_{\mathrm{mask}}(\theta,s)}\right)
            =
            \int_{\ell(\theta,s)} c(x,y,t)\,dl.
            \label{eq:conc_to_sinogram}
        \end{equation}
        
\begin{figure}[htbp]
            \centering
            \includegraphics[scale=0.53,trim={0 0 0 0},clip]{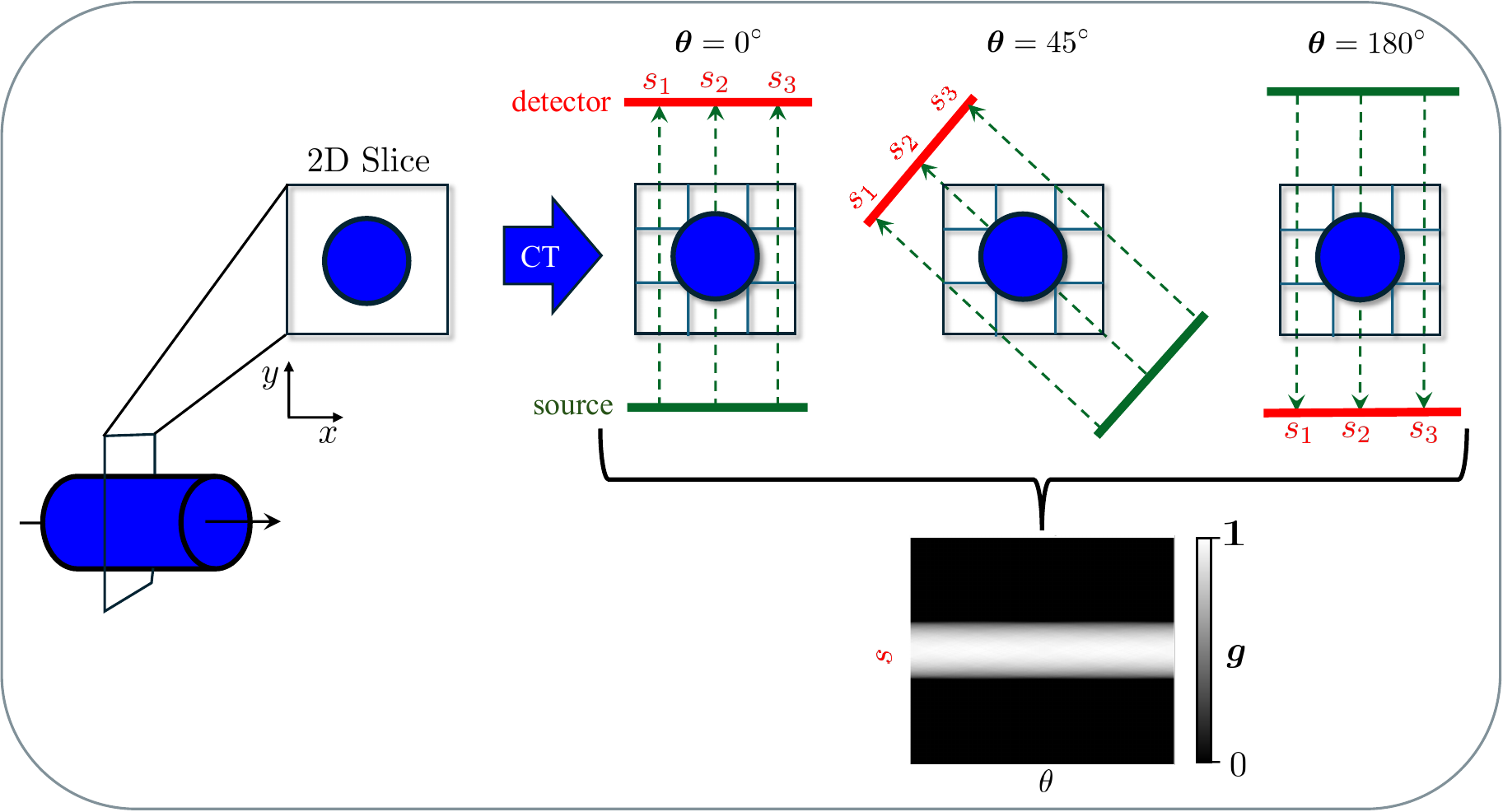}
            \caption{Schematic illustration of parallel-beam CT forward projection and sinogram generation. A time-resolved contrast concentration field on the discretized domain is forward-projected at multiple angles, and the resulting line-integrated detector measurements are assembled into the sinogram.}
            \label{fig:sinogram_generation}
        \end{figure}

        In the present formulation, the transport time index and projection-angle index are treated as independent coordinates for the projection. The instantaneous concentration snapshot $c(x, y,t_k)$ is forward-projected over the prescribed set of projection angles $\theta\in[0,\pi)$ at each transport time level $t_k$ using \cref{eq:conc_to_sinogram}. This produces a time-resolved predicted sinogram sequence $\hat{g}_{t_k,s,\theta}(\boldsymbol{\gamma})$. Sparse-view acquisition is then modeled by evaluating the data-fidelity loss only on the available subset of projection angles $\Theta_{\mathrm{sparse}}$. \Cref{fig:sinogram_generation} illustrates the projection geometry and the assembly of detector measurements into the sinogram. In this implementation, the forward model evaluates a fixed set of projection angles at each discrete time step. Future work will extend this framework to simulate a conventional rotating gantry, where the projection angle and acquisition time are coupled. 
        
        \subsection{Nonlinear Solver}
            \label{subsec:fem_formulation}
            With the material representation established and the effective physical properties computed for each element based on the design field, the system response is evaluated by solving the governing equations detailed in \cref{subsec:governing_equations} using the finite element method.

    \subsubsection*{Steady-State Navier--Stokes Discretization}
         We discretize the strong-form residual $\mathcal{R}_{\mathrm{NS}}$ of the incompressible, steady-state Navier--Stokes equations (\cref{eq:navier_stokes_system}) to obtain a nonlinear algebraic system for the nodal velocity--pressure vector $\mathbf{U}=[\mathbf{u},\mathbf{p}]^T$:
        \begin{equation}
            \mathbf{R}_{\mathrm{NS}}(\mathbf{U},\boldsymbol{\gamma})
            =
            \begin{bmatrix}
                \mathbf{R}_m(\mathbf{u},\mathbf{p},\boldsymbol{\gamma}) \\
                \mathbf{R}_c(\mathbf{u})
            \end{bmatrix}
            =
            \mathbf{0},
            \label{eq:ns_discrete_residual}
        \end{equation}

        where $\mathbf{R}_m$ and $\mathbf{R}_c$ denote the discrete momentum and continuity residuals, respectively. Note that $\boldsymbol{\gamma}$ appears in the discrete system through the interpolated Brinkman penalty coefficient $\alpha(\boldsymbol{\gamma})$, as detailed in \cref{subsec:design_representation}. In matrix form, the momentum residual is assembled from the convective, viscous, pressure-gradient, and Brinkman-penalization contributions:

            \begin{equation}
                \mathbf{R}_m(\mathbf{u},\mathbf{p},\boldsymbol{\gamma})
                =
                \mathbf{C}(\mathbf{u})\mathbf{u}
                +
                \mathbf{K}\mathbf{u}
                +
                \mathbf{B}^T\mathbf{p}
                +
                \mathbf{A}_{\alpha}(\boldsymbol{\gamma})\mathbf{u}
                -
                \mathbf{b},
                \label{eq:ns_momentum_matrix}
            \end{equation}
            while the incompressibility is enforced through the continuity residual:

            \begin{equation}
                \mathbf{R}_c(\mathbf{u})=\mathbf{B}\mathbf{u}.
                \label{eq:ns_continuity_matrix}
            \end{equation}
            
        Here, $\mathbf{C}(\mathbf{u})$ is the nonlinear advection matrix, $\mathbf{K}$ is the viscous stiffness matrix, $\mathbf{B}$ is the discrete divergence matrix, $\mathbf{B}^T$ is the pressure-gradient matrix, $\mathbf{A}_{\alpha}(\boldsymbol{\gamma})$ is the design-dependent Brinkman penalization matrix, and $\mathbf{b}$ is the boundary condition contribution vector. Using the velocity shape functions $\mathbf{N}_u$ and the pressure shape functions $\mathbf{N}_p$, these matrices are assembled from their element-level contributions:
        
        \begin{subequations}\label{eq:ns_elemental_matrices}
            \begin{align}
                \mathbf{C}^{(e)}(\mathbf{u}) &= \int_{\Omega_e} \mathbf{N}_u^T \rho \left(\mathbf{u} \cdot \nabla \mathbf{N}_u\right) \,\mathrm{d}\Omega, 
                \label{eq:ns_elemental_advection} \\
                \mathbf{K}^{(e)} &= \int_{\Omega_e} \nabla \mathbf{N}_u^T \mu \left( \nabla \mathbf{N}_u + \nabla \mathbf{N}_u^T \right) \,\mathrm{d}\Omega, 
                \label{eq:ns_elemental_viscous} \\
                \mathbf{B}^{(e)} &= \int_{\Omega_e} \mathbf{N}_p^T \left(\nabla \cdot \mathbf{N}_u\right) \,\mathrm{d}\Omega, 
                \label{eq:ns_elemental_divergence} \\
                \mathbf{A}_{\alpha}^{(e)}(\gamma_e) &= \int_{\Omega_e} \mathbf{N}_u^T \alpha(\gamma_e) \mathbf{N}_u \,\mathrm{d}\Omega.
                \label{eq:ns_elemental_brinkman}
            \end{align}
        \end{subequations}
        
        To circumvent the Ladyzhenskaya--Babuška--Brezzi (LBB) condition and permit equal-order interpolation for velocity and pressure, the standard Galerkin formulation is augmented with PSPG and SUPG stabilization to suppress spurious pressure oscillations and convective instabilities, respectively \cite{tezduyar2001adaptive, alexandersen2023detailed}. For more details on stabilization terms, the reader is referred to \cite{tezduyar2001adaptive, alexandersen2023detailed}.
            
   \subsubsection*{Transient Transport Discretization}

        Having computed the steady-state velocity field $\mathbf{u}$, we evaluate the transient concentration response by discretizing \cref{eq:transient_transport_residual}. The domain is discretized using four-noded bilinear quadrilateral elements with shape functions $\mathbf{N}$; for temporal discretization, we employ the implicit first-order backward-Euler scheme. This yields the discrete algebraic system for the nodal concentration vector $\mathbf{c}^{n+1}$ at time step $n+1$:
        
        \begin{equation}
            \left(\mathbf{M} + \Delta t\,\mathbf{C}_{t} + \Delta t\,\mathbf{K}_t\right)\mathbf{c}^{n+1}
            =
            \mathbf{M}\mathbf{c}^{n} + \Delta t\,\mathbf{b}^{n+1},
            \label{eq:implicit_euler_discretization}
        \end{equation}
        
        where $\Delta t$ is the time-step size and $\mathbf{b}^{n+1}$ denotes the boundary condition contribution at time step $n+1$. The global mass matrix $\mathbf{M}$, advection matrix $\mathbf{C}_t$, and design-dependent diffusion matrix $\mathbf{K}_t$ are assembled from their element-level contributions:
        
        \begin{subequations}\label{eq:elemental_matrices}
            \begin{align}
                \mathbf{M}^{(e)} &= \int_{\Omega_e} \mathbf{N}^T \mathbf{N}\,\mathrm{d}\Omega, 
                \label{eq:elemental_mass_matrix} \\
                \mathbf{C}_t^{(e)} &= \int_{\Omega_e} \mathbf{N}^T \left(\mathbf{u}\cdot\nabla \mathbf{N}\right)\,\mathrm{d}\Omega, 
                \label{eq:elemental_advection_matrix} \\
                \mathbf{K}_{t}^{(e)}({\gamma}_e) &= \int_{\Omega_e} \nabla \mathbf{N}^T
                D({\gamma}_e)\,
                \nabla \mathbf{N}\,\mathrm{d}\Omega.
                \label{eq:elemental_diffusion_matrix}
            \end{align}
        \end{subequations}
        Note that $\boldsymbol{\gamma}$ appears in the discrete system through the interpolated mass diffusivity $D(\gamma_e)$ via the design-dependent diffusion matrix $\mathbf{K}_{t}^{(e)}(\gamma_e)$, as detailed in \cref{subsec:design_representation}. Furthermore, to mitigate spurious oscillations characteristic of convection-dominated transport, the standard Galerkin formulation is augmented with SUPG stabilization \cite{tezduyar2001adaptive, noorishad1994streamline}.

        The steady flow system is solved by a modified Newton--Raphson iterative scheme. The transport system is linear in the concentration, so each time step requires a single linear solve. All linear systems are solved with PETSc \cite{balay2026petsc, dalcin2011parallel}. The entire simulation pipeline is implemented within the JAX ecosystem \cite{bradbury2018jax}, where the iterative solvers are wrapped in custom Jacobian-vector products, rendering the framework end-to-end differentiable. This enables exact and efficient design-sensitivity computation as discussed next.

        \subsection{Sensitivity Analysis}
        \label{subsec:sensitivity_analysis}

            A critical step in gradient-based TO is the exact computation of sensitivities, the derivatives of the objective function with respect to the design variables. The sensitivity analysis for this work is particularly complex, as it must account for the coupled Navier--Stokes and transient advection-diffusion forward models, as well as the tomographic projection operator. To address this complexity, we construct an end-to-end differentiable pipeline using the automatic differentiation (AD) capabilities of the JAX framework \cite{bradbury2018jax}. This allows us to avoid the laborious and error-prone process of manually deriving adjoint sensitivity expressions \cite{padhy2026toflux, padhy2026moto}. By implementing the entire forward analysis within JAX, from the material interpolation to the final tomographic projection, the framework computes the required design sensitivities to machine precision via reverse-mode AD.

            We emphasize two AD challenges inherent to our multiphysics simulation. The first arises from the use of iterative Newton--Raphson solvers to resolve the steady fluid fields. A naive application of reverse-mode AD would unroll derivative computation across every solver iteration, computing the total derivative via the chain rule:

             \begin{equation}
                \frac{\mathrm{d}\mathbf{U}^{(K)}}{\mathrm{d}\gamma}
                =
                \frac{\partial \mathbf{U}^{(K)}}{\partial \mathbf{U}^{(K-1)}}
                \cdot
                \frac{\partial \mathbf{U}^{(K-1)}}{\partial \mathbf{U}^{(K-2)}}
                \cdots
                \frac{\partial \mathbf{U}^{(1)}}{\partial \mathbf{U}^{(0)}}
                \cdot
                \frac{\partial \mathbf{U}^{(0)}}{\partial \gamma}.
                \label{eq:loop_unroll}
            \end{equation}
            
            This approach is computationally expensive and memory-intensive, as both scale with the number of iterations $K$. To overcome this inefficiency, we apply the Implicit Function Theorem (IFT) \cite{blondel2022efficient}. Given that the residual vanishes at convergence, $\mathbf{R}_{\mathrm{NS}}(\gamma, \mathbf{U}^{(K)}) = \mathbf{0}$, and the Jacobian $\partial \mathbf{R}_{\mathrm{NS}}/\partial \mathbf{U}$ is invertible, the derivative follows directly:
            
            \begin{equation}
                \frac{\mathrm{d}\mathbf{U}^{(K)}}{\mathrm{d}\gamma}
                =
                -\left(\frac{\partial \mathbf{R}_{\mathrm{NS}}}{\partial \mathbf{U}}\right)^{-1}
                \frac{\partial \mathbf{R}_{\mathrm{NS}}}{\partial \gamma}.
                \label{eq:ift}
            \end{equation}
            
            This enables direct computation of sensitivities from the final converged solution, bypassing the need to backpropagate through the iterative solution history. 

            The second challenge stems from the transient nature of the contrast transport. Reverse-mode AD through the time loop requires the forward concentration state at every time step, because the gradient is accumulated by solving one transposed linear system per step, backward in time.  In the present work, the full state history is stored for the backward pass. Its memory grows linearly with the number of time steps, but remains a small fraction of the memory of one objective and gradient evaluation, which is dominated by the forward projection and the loss. When the stored history becomes too large, for example on finer meshes or over longer simulated times, \emph{checkpointing} can be implemented to reduce the memory by storing only selected states and recomputing the others during the backward pass \cite{padhy2026tomatoes, wang2009minimal, james2015topology}.

    \subsection{Optimization}
    \label{subsec:optimization}
            
            We now formulate the inverse problem as a PDE-constrained optimization task. The key components of the optimization framework are outlined below.
            
            \textbf{Design Variables:} The optimization contains two sets of design variables. The first is the pseudo-density field $\boldsymbol{\gamma} \in \mathbb{R}^{n_e}$, where $n_e$ is the number of mesh elements. $\gamma_e$ determines whether the element behaves as fluid or solid through the material interpolation described in \cref{subsec:design_representation}. The second set consists of the inlet velocity scaling parameters $\mathbf{a}=[a_1,\ldots,a_{n_{\mathrm{in}}}]^\top$, where $n_{\mathrm{in}}$ is the number of inlet boundaries.  Specifically, the velocity profile on inlet $i$ is written as $\mathbf{u}_{\mathrm{in},i} = a_i U_{\mathrm{in}}(y), \qquad i=1,\ldots,n_{\mathrm{in}}$, where $U_{\mathrm{in}}(y)$ is the nominal parabolic inlet profile and $a_i$ is a dimensionless scaling factor (\cref{fig:bifurcated_geom_gt_op}(a)). The scaling parameters are critical for optimizing the inlet flow conditions depending on downstream constraints.

            \textbf{Objective:} The primary goal is to minimize the discrepancy between the simulated tomographic projections and the measured synthetic data. To reflect sparse-view CT acquisition conditions, this loss is evaluated at the projection angles, denoted by the set $\Theta_{\mathrm{sparse}}$. Specifically, the objective function $J(\boldsymbol{\gamma}, \mathbf{a})$ is defined as the Mean Squared Error (MSE) in the measurement space:
            \begin{equation}
                J(\boldsymbol{\gamma}, \mathbf{a})
                =
                \frac{1}{N_t N_s |\Theta_{\mathrm{sparse}}|}
                \sum_{k=1}^{N_t}
                \sum_{s=1}^{N_s}
                \sum_{\theta \in \Theta_{\mathrm{sparse}}}
                \left(
                \hat{g}_{t_k,s,\theta}(\boldsymbol{\gamma}, \mathbf{a}) - g^{\mathrm{data}}_{t_k,s,\theta}
                \right)^2,
                \label{eq:loss_function}
            \end{equation}
            where $N_t$ and $N_s$ denote the number of frames and the number of detector positions, respectively, and $|\Theta_{\mathrm{sparse}}|$ denotes the number of measured projection angles. Here, $\hat{g}_{t_k,s,\theta}(\boldsymbol{\gamma}, \mathbf{a})$ is the predicted transient sinogram obtained by applying the Radon transform to the simulated concentration field, while $g^{\mathrm{data}}_{t_k,s,\theta}$ denotes the corresponding target sinogram data acquired from the CE-CT scan. 
            
            \textbf{Bound Constraints:} The pseudo-density variables are constrained by $\gamma_e \in [0, 1]$, where $0$ denotes the fluid region (vascular lumen) and $1$ denotes the solid region (surrounding tissue). The inlet scaling parameters are also bounded to keep the imposed flow rates within a prescribed admissible range, $a_i^{\min} \leq a_i \leq a_i^{\max}, \qquad i=1,\ldots,n_{\mathrm{in}}$
            
            \textbf{Optimization:} Collecting the objective (\cref{eq:loss_function}), the governing steady flow and transient transport state residuals (\cref{eq:navier_stokes_system}, \cref{eq:transient_transport_residual}), and the bound constraints, the continuous optimization problem can be expressed as:
            \begin{subequations}\label{eq:optimization_problem}
                \begin{align}
                \min_{\boldsymbol{\gamma}, \mathbf{a}} \quad & J(\boldsymbol{\gamma}, \mathbf{a}) \\
                \text{subject to} \quad &
                \mathcal{R}_{\mathrm{NS}}\big(\mathbf{u}, \mathbf{p} \big)=0
                \qquad \text{in } \Omega, \\
                &
                \mathcal{R}_{\mathrm{T}}\big( \mathbf{c}, \mathbf{u} \big)=0
                \qquad \text{in } \Omega \times (0,T], \\
                &
                0 \le \gamma_e \le 1, \qquad e=1,\dots,n_e, \\
                &
                a_i^{\min} \leq a_i \leq a_i^{\max},
                \qquad i=1,\ldots,n_{\mathrm{in}}.
                \end{align}
            \end{subequations}
            
            As outlined in the preceding sections, the design variables dictate the Brinkman penalization and effective diffusivity, which govern the steady flow and transient transport solutions, respectively. These coupled physical states are ultimately forward-projected to generate the predicted sinograms $\hat{g}_{t_k,s,\theta}(\boldsymbol{\gamma}, \mathbf{a})$.
                        
            To navigate the design landscape, the optimization is carried out using the Method of Moving Asymptotes (MMA) \cite{svanberg1987method}. The exact gradients of the objective with respect to the design variables are evaluated efficiently using reverse-mode automatic differentiation through the full PDE-constrained computational pipeline, as detailed in \cref{subsec:sensitivity_analysis}. Finally, to prevent convergence to poor local minima and promote a well-posed, purely binary fluid-solid design, continuation strategies are employed. Specifically, the Brinkman penalization parameter ($q_\alpha$), diffusivity parameter ($q_D$) and the steepness of the Heaviside projection function are incrementally updated during the optimization process.

\section{Numerical Experiments}
\label{sec:numerical_experiments}

    To evaluate the proposed physics-constrained topology reconstruction framework, we present a series of numerical experiments. To quantitatively assess the accuracy of the continuous-valued reconstruction, we calculate the normalized root mean square error (nRMSE) between the optimized pseudo-density field $\boldsymbol{\gamma}$ and the ground-truth geometry $\boldsymbol{\gamma}_{\mathrm{true}}$ over the $N_e$ elements \cite{yoon2010simultaneous, romanov2016simultaneous}. 
    Furthermore, to evaluate the structural accuracy of the final thresholded designs, we utilize the Dice similarity coefficient (DSC). The DSC is widely established in the tomographic and medical imaging literature for quantifying spatial overlap and boundary agreement between a reconstructed binary shape and its corresponding ground truth \cite{sadid2022segmenting, wang2023automatic, huang2022simultaneous}. The reconstructed flow is compared with the true flow, through the relative velocity error, $\varepsilon_{\mathbf{u}}$ over the true lumen.

    All experiments were conducted on a MacBook M4 Pro, using the JAX \cite{bradbury2018jax} library in Python. Unless otherwise specified, the default parameters for all numerical examples are as follows:

    \subsection{Default Parameters}
    \label{subsec:numerical_experiments_default}
    
    Unless otherwise stated, all experiments use the parameters in \cref{tab:default_parameters}; geometry-specific boundary conditions and inlet scalings are given with each example. The paragraphs below define the quantities in the table where needed and explain the choices that are not standard.
    
    \begin{table}[htbp]
        \centering
        \caption{Default parameters. CGS units, with SI equivalents where helpful.}
        \label{tab:default_parameters}
        \begin{tabular}{lll}
            \hline
            Quantity & Symbol & Value \\
            \hline
            \multicolumn{3}{l}{\textbf{Discretization and design representation}} \\
            \hspace{1em}Domain & $\Omega$ & $1\times1\,\mathrm{cm}$ \\
            \hspace{1em}Mesh & -- & $80\times80$ bilinear quadrilaterals\\
            \hspace{1em}Filter radius & $r$ & $5.5\%$ of the domain diagonal \\
            \hspace{1em}Heaviside threshold & $\eta$ & $0.5$ \\
            \hspace{1em}Initial design & $\gamma_e^{(0)}$ & $0.5$ \\
            \multicolumn{3}{l}{\textbf{Flow}} \\
            \hspace{1em}Density & $\rho$ & $1.058\,\mathrm{g/cm^3}$ ($1058\,\mathrm{kg/m^3}$) \\
            \hspace{1em}Viscosity & $\mu$ & $0.0345\,\mathrm{P}$ ($3.45\,\mathrm{mPa\,s}$) \\
            \hspace{1em}Inlet width & $w$ & $0.15\,\mathrm{cm}$ \\
            \hspace{1em}Reynolds number & $Re$ & $7.68$ \\
            \hspace{1em}Brinkman bounds & $\alpha_{\min},\,\alpha_{\max}$ & $8.6\times10^{-6},\ 3.45\times10^{4}\,\mathrm{g\,cm^{-3}s^{-1}}$ \\
            \multicolumn{3}{l}{\textbf{Transport}} \\
            \hspace{1em}Diffusivity (fluid, solid) & $D_{\max},\,D_{\min}$ & $5\times10^{-3},\ 5\times10^{-6}\,\mathrm{cm^2/s}$ \\
            \hspace{1em}Simulated time; time step & $T$; $\Delta t$ & $3\,\mathrm{s}$; $1/320\,\mathrm{s}$ \\
            \hspace{1em}Injection & $c_{\mathrm{in}}=1$ & $t\in[0.025,0.525]\cup[1.525,2.025]\,\mathrm{s}$ \\
            \multicolumn{3}{l}{\textbf{Projection}} \\
            \hspace{1em}Frames & $N_t$ & 121 \\
            \hspace{1em}Detector bins; angles & $N_s$; $N_\theta$ & 115; 180 over $[0,\pi)$ \\
            \hspace{1em}Angles in the loss & $|\Theta_{\mathrm{sparse}}|$ & 90 (every second angle) \\
            \multicolumn{3}{l}{\textbf{Solvers and optimization}} \\
            \hspace{1em}Newton-Raphson (flow) & -- & modified; $\|\mathbf{r}\|\le10^{-8}\|\mathbf{r}_0\|$, $\le10$ iterations \\
            \hspace{1em}Linear solver & -- & PETSc, rtol $10^{-10}$ \\
            \hspace{1em}Optimizer & -- & MMA, move limit $0.05$, 300 iterations \\
            \hspace{1em}Continuation, \cref{eq:ramp_brinkman} & $q_\alpha$ & $75\to50$, $-0.25$ per iteration \\
            \hspace{1em}Continuation, Heaviside & $\beta$ & $1\to4$, $+0.05$ per iteration \\
            \hspace{1em}Continuation, \cref{eq:ramp_diffusivity} & $q_D$ & $5\to10$, $+0.05$ per iteration \\
            \hline
        \end{tabular}
    \end{table}

    \paragraph{Flow.}
    The Brinkman bound $\alpha_{\max}$ is chosen relative to the inlet width length scale. It corresponds to a Darcy number $Da=\mu/(\alpha_{\max}w^2)=4.4\times10^{-5}$, close to the value used by \cite{alexandersen2023detailed} for Navier--Stokes channel flow with an inlet-width-based Reynolds number. The accuracy of this choice is verified in \cref{subsection:numerical_parameters}.
    
    \paragraph{Transport.}
    The fluid diffusivity $D_{\max}$ represents an effective dispersion coefficient. The transport equation is integrated with the backward Euler scheme (960 steps). The contrast agent is injected through all inlets in two boluses; each switch of $c_{\mathrm{in}}$ is smoothed by a raised-cosine ramp of width $0.05\,\mathrm{s}$, and $c_{\mathrm{in}}=0$ outside the injection intervals.
    
    \paragraph{Projection.}
    The 121 frames are taken every $0.025\,\mathrm{s}$, including $t=0$, and each ray is integrated with 115 samples.
    
    \paragraph{Solvers and optimization.}
    The modified Newton--Raphson method follows \cite{alexandersen2023detailed}. Because the transport equation is linear in $c$ for a given velocity field, its system matrix is assembled once per design, and each time step requires a single linear solve. The design is updated with MMA \cite{svanberg1987method}. Because the objective decreases by several orders of magnitude during the optimization, it is divided by its current value whenever it has fallen by a factor of five since the last rescaling \cite{suresh2021design}; this keeps the gradients well above the internal regularization of MMA. Each flow solve converges in three to five Newton iterations (three in most solves). The gradient was checked against central finite differences.
    
    \subsection{Synthetic Data Generation and Sparse-View Measurements}
    \label{subsec:synthetic_data_generation}
            
        We evaluate the framework on synthetic data so that the reconstructed geometry and flow field can be scored against the true, known vessel geometry and flow field. For each ground-truth geometry, the data is produced with the forward model of \cref{subsec:governing_equations,subsec:fwd_projection}: the steady flow is solved on the ground-truth design, the concentration transport is driven by the resulting flow field, and the concentration field is projected at every frame. The true geometry and velocity fields are stored for evaluation and are never used by the reconstruction.

        To avoid an inverse crime \cite{sixou2020contrast}, the ground-truth data is generated with a forward model with a finer discretization and it differs from the one used in the reconstruction, as summarized in \cref{tab:data_vs_reconstruction}. 

        \begin{table}[htbp]
            \centering
            \caption{Forward model used to generate the synthetic data and forward model used in the reconstruction.}
            \label{tab:data_vs_reconstruction}
            \begin{tabular}{lll}
                \hline
                & Data generation & Reconstruction \\
                \hline
                Mesh & $160\times160$ & $80\times80$ \\
                Time step $\Delta t$ & $1/1280\,\mathrm{s}$ (3840 steps) & $1/320\,\mathrm{s}$ (960 steps) \\
                Samples per ray & 912 & 115 \\
                projections & 180 & 90 \\
                Physical parameters & as in \cref{subsec:numerical_experiments_default} & as in \cref{subsec:numerical_experiments_default} \\
                \hline
            \end{tabular}
        \end{table}

        Three ground-truth geometries are considered for the framework evaluation. First an idealized bifurcated artery, in which two inflow branches merge. Second, an idealized stenosed channel with $50\%$ lumen narrowing and third, an idealized carotid bifurcation with an asymmetric stenosis in one branch. All three use the transport protocol of \cref{subsec:numerical_experiments_default}; their boundary conditions are given in the corresponding subsections. Each ground-truth geometry is defined by a smooth analytic or drawn boundary and rasterized directly at the required resolution, so no geometry is interpolated between meshes. The ground-truth design is passed through the same density filter and Heaviside projection as the reconstruction, so the two share the same design representation.

        Because the synthetic data are generated with a finer discretization, the reconstruction model cannot reproduce the data exactly, even for the correct geometry. To measure this error, we take the ground-truth geometry and compute its sinograms with the reconstruction model. These sinograms differ from the synthetic data by $1.2\%$ in the relative $L_2$ norm. We call this difference the model error.

    \subsection{Joint Reconstruction of Vascular Topology and Inlet Flow Parameters} 
    \label{subsec:bifurcated_artery}

\begin{figure}[htbp]
         	\centering
        		\includegraphics[scale=0.55,trim={0 0 0 0},clip]{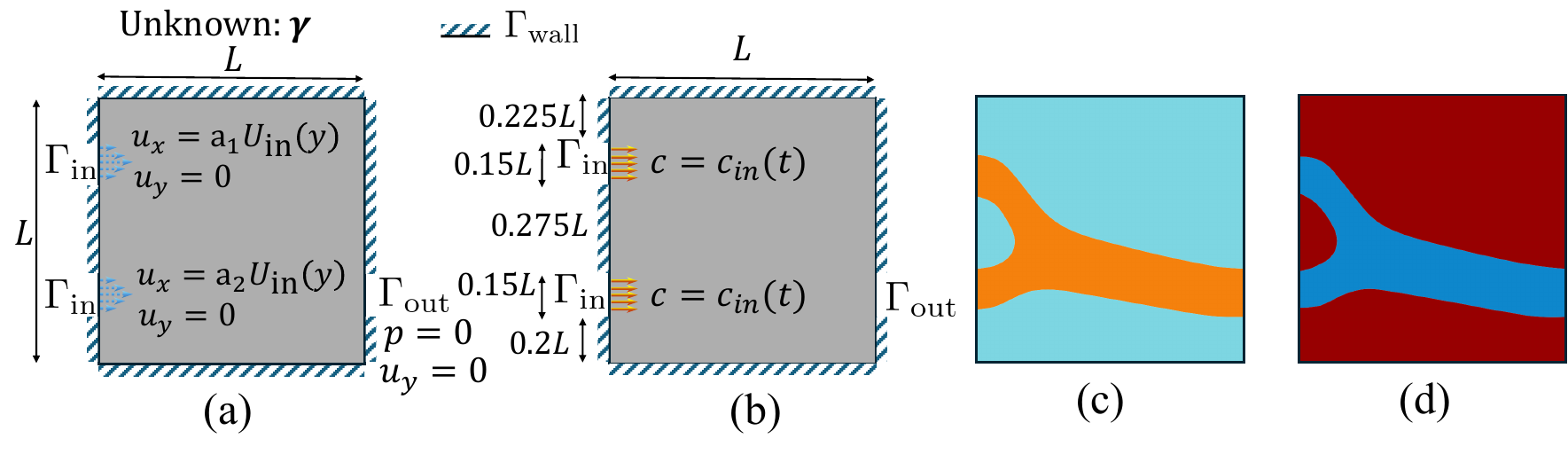}
         		\caption{Joint topology and inlet-flow reconstruction setup.
                (a) Fluid flow boundary conditions. 
                (b) Transient transport boundary conditions. 
                (c) Ground-truth geometry used to generate the synthetic data. 
                (d) Recovered binary geometry obtained by jointly optimizing the pseudo-density field $\boldsymbol{\gamma}$ and inlet scaling parameters $\mathbf{a}$ from sparse time-resolved sinogram data.
                }
                \label{fig:bifurcated_geom_gt_op}
         \end{figure}

        In this experiment, we simultaneously recover both an unknown vessel geometry and its unknown inlet flow scalings from sinograms. The ground truth is an idealized bifurcated artery in which two inflow branches merge into a single outlet (\cref{fig:bifurcated_geom_gt_op}(c)). The flow and transport boundary conditions are shown in \cref{fig:bifurcated_geom_gt_op}(a-b). The setup is shown in \cref{fig:bifurcated_geom_gt_op}. Parabolic profiles with scaling parameters $a_1$ (upper inlet) and $a_2$ (lower inlet) are prescribed at the inlets, and both are unknown in the reconstruction together with the design field $\boldsymbol{\gamma}$. The data are the time-resolved sinograms, generated as described in \cref{subsec:synthetic_data_generation}.

        Starting from a uniform initial design and low inlet flow scalings  $\mathbf{a}^{(0)}\approx[0.6, 0.6]^\top$, the reconstruction recovers the bifurcation, the junction and both inflow rates (\cref{fig:bifurcated_geom_gt_op}(d)). The Dice coefficient is $\mathrm{DSC}=0.993$. The recovered inlet scalings are $\mathbf{a}^\star=[1.005,\,0.996]^\top$, within $0.5\%$ of their true values. The velocity error over the true lumen is $\varepsilon_{\mathbf{u}}=2.3\%$. These errors are of the same order as the Brinkman modeling error of the reconstruction model itself (\cref{subsection:numerical_parameters}), so they are not dominated by errors in the recovered geometry. 
        
\begin{figure}[htbp]
            \centering
                \includegraphics[scale=0.5,trim={0 0 0 0},clip]{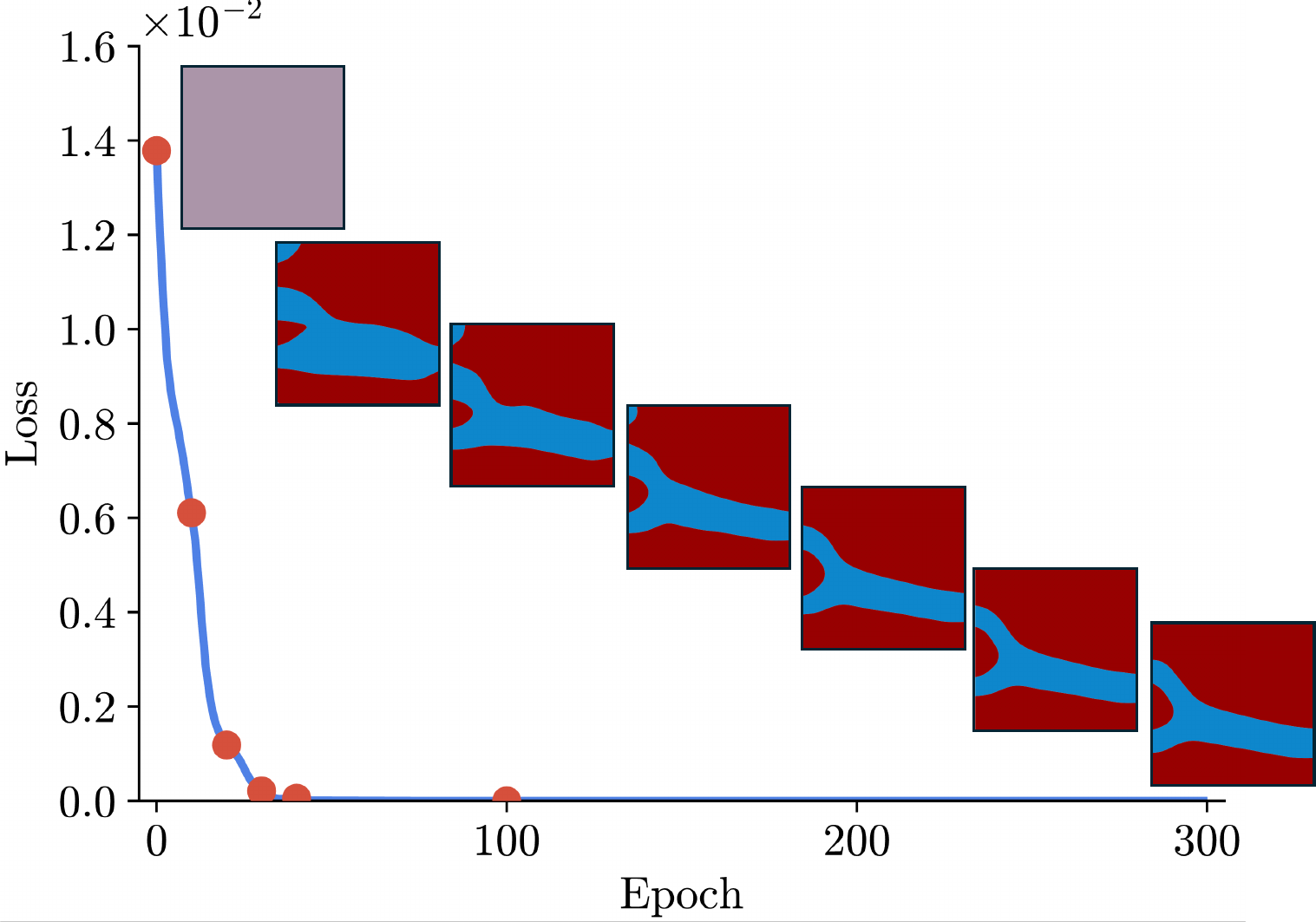}
                \caption{ Bifurcated-artery benchmark: convergence of the objective.}
                \label{fig:bifurcated_convergence}
         \end{figure}

          \Cref{fig:bifurcated_convergence} shows how the design evolves. The objective decreases steeply while the continuation parameters change (iterations 1--100), and the topology is established early: the separating wall between the two inflow branches reaches its true length by iteration 34. After continuation ends, the objective oscillates between two nearby values on alternate iterations (by about 1\% on the $80\times80$ mesh), while the design no longer changes appreciably: the Dice coefficient changes by less than $0.001$ between iteration 124 and iteration 300. One iteration takes $6.8\,\mathrm{s}$ on the $80\times80$ mesh (about $34\,\mathrm{min}$ for 300 iterations) with a peak memory of $5.3\,\mathrm{GB}$; the scaling with mesh size is reported in \cref{subsection:mesh_dependency}.
          
\begin{figure}[htbp]
            \centering
                \includegraphics[scale=0.55,trim={0 0 0 0},clip]{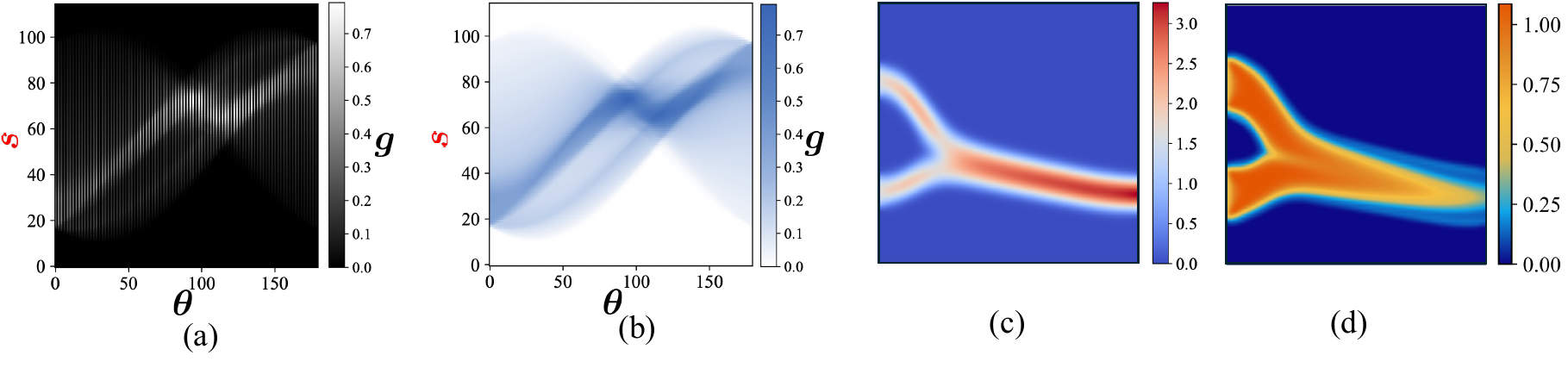}
                \caption{ Bifurcated-artery benchmark.  
                (a) Sparse sinogram frame at $t = 2.025\,\mathrm{s}$ generated from ground-truth geometry for synthetic data. 
                (b) Predicted sinogram frame generated from the concentration field at the same time. 
                (c) Predicted velocity-magnitude field. 
                (d) Predicted contrast concentration field at the same time. 
                }
                \label{fig:bifurcated_estg_estvel}
         \end{figure}

        The predicted sinogram in \cref{fig:bifurcated_estg_estvel}(b), obtained by projecting the optimized concentration field, matches the sparse target in \cref{fig:bifurcated_estg_estvel}(a). The associated velocity and concentration fields at $t=2.025\,\mathrm{s}$ through the bifurcated lumen are shown in \cref{fig:bifurcated_estg_estvel}(c,d). The geometry and the inlet-flow scalings are thus recovered through the flow and transport physics, not by an image-based segmentation.

    \subsection{Mesh Dependency}
    \label{subsection:mesh_dependency}

        Density-based topology optimization is known to produce mesh-dependent designs unless the design is regularized with a length scale \cite{sigmund1998numerical}. We therefore test whether the reconstruction converges as the mesh is refined, with the filter radius held fixed in physical units. The bifurcation geometry is reconstructed on the three meshes from the same data, generated on the $160\times160$ mesh; all other settings follow \cref{tab:default_parameters}.
        
        \begin{table}[htbp]
            \centering
            \caption{Bifurcation reconstructed on three meshes from the same $160\times160$ data. Time per iteration measured for a single run.}
            \label{tab:mesh_study}
            \begin{tabular}{lccc}
                \hline
                Mesh & $40\times40$ & $80\times80$ & $120\times120$ \\
                \hline
                $\mathrm{DSC}$ & 0.986 & 0.993 & 0.996 \\
                $\mathrm{nRMSE}_\gamma$ & 0.026 & 0.016 & 0.011 \\
                $\varepsilon_{\mathbf{u}}$ [\%] & 4.0 & 2.3 & 1.9 \\
                Inlet scaling $a_1^\star$ / $a_2^\star$ & $1.030,\ 1.010$ & $1.005,\ 0.996$ & $1.000,\ 0.999$ \\
                Time per iteration [s] & 2.2 & 6.8 & 18.5 \\
                \hline
            \end{tabular}
        \end{table}
        
\begin{figure}[htbp]
            \centering
            \includegraphics[scale=0.35,trim={0 0 0 0},clip]{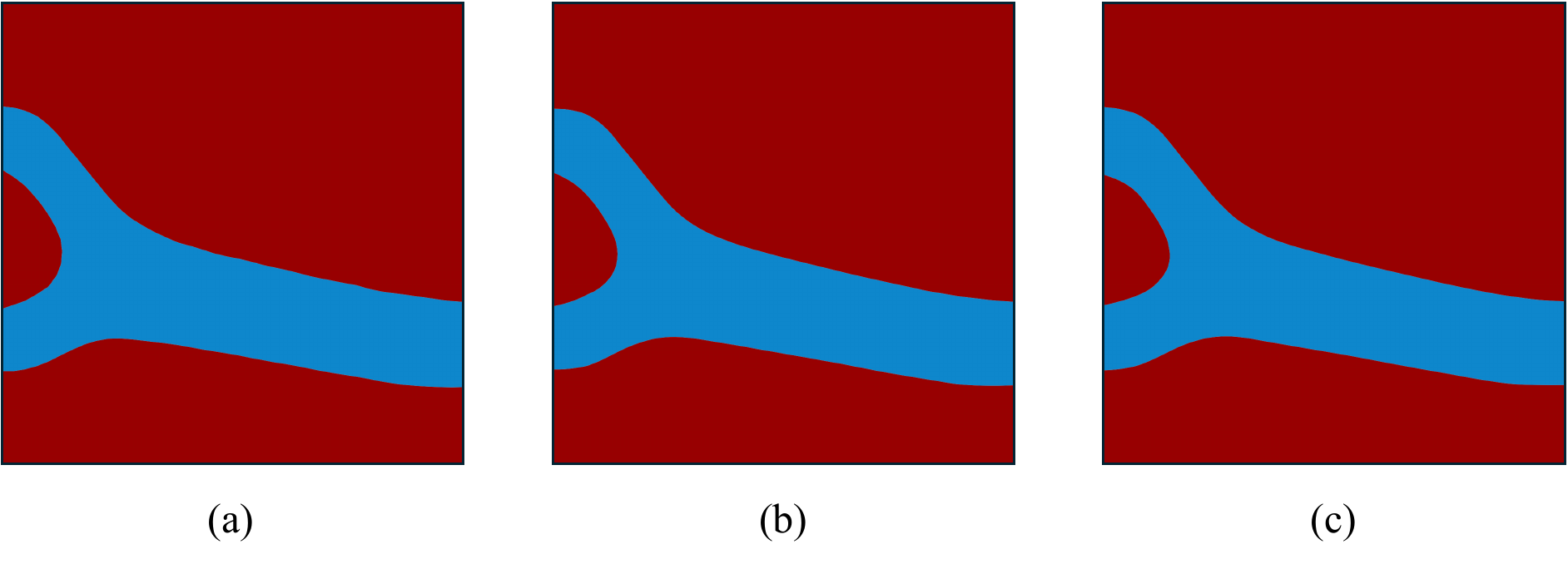}
            \caption{Reconstructions of the bifurcation on (a) $40\times40$, (b) $80\times80$ and (c) $120\times120$ meshes. $\mathrm{DSC}=0.986$, $0.993$, $0.996$.}
            \label{fig:mesh_study}
        \end{figure}

        The reconstruction converges with the mesh (\cref{tab:mesh_study,fig:mesh_study}). The topology is the same on all three meshes: two inflow branches merging into one outlet, with the separating wall in the same place. Refinement improves only the wall position. The DSC rises from $0.986$ to $0.996$. The design-field error decreases from $0.026$ to $0.011$, and the velocity error from $4.0\%$ to $1.9\%$. The inlet scalings are recovered within $3\%$, $0.5\%$ and $0.1\%$ on the three meshes, respectively.
        
        The computational cost grows with the mesh. The time per iteration grows from $2.2\,\mathrm{s}$ to $18.5\,\mathrm{s}$ as the number of elements increases ninefold, roughly in proportion to the number of degrees of freedom, so that the 300 iterations take about 11, 34 and 93 minutes on the three meshes, respectively. The peak memory grows from $2.7$ to $5.3$ and to $7.9\,\mathrm{GB}$ with the mesh. The transient transport accounts for about 83\% of the time on every mesh, split almost equally between the forward solve and its reverse-mode differentiation, because each of the 960 time steps requires one linear solve in each pass; the flow solve takes 5--11\% and the projection 3--12\%.
        
         The time step is the same on all meshes ($\Delta t=1/320\,\mathrm{s}$), so the Courant number on the $120\times120$ mesh is $1.26$, compared with $0.85$ on the $80\times80$ mesh. The implicit scheme remains stable, but the temporal error is larger. A forward test on the true geometry shows that halving the time step on the $120\times120$ mesh reduces the model error with respect to the data from $0.91\%$ to $0.40\%$. The accuracy reported for the $120\times120$ mesh is therefore conservative.
        
    \subsection{Robustness to model mismatch}
    \label{subsection:model_mismatch}
            
        The reconstruction assumes that the physical model is known: the blood viscosity $\mu$, the contrast diffusivity $D$, and the locations of the inlets and outlets. In practice, the viscosity and the contrast dispersion of a patient are not measured, and literature values are assumed \cite{nolte2022inverse}. The boundary locations are taken from the image and carry an error of the order of the image resolution. We therefore ask how much the reconstruction degrades when the assumed model differs from the model that produced the data.
        
        We test three mismatches, one at a time. In each test, a single quantity differs between the data and the reconstruction; all other settings are those of \cref{subsec:numerical_experiments_default}, and the true geometry is the bifurcated artery (\cref{fig:bifurcated_geom_gt_op}(c)).
        \begin{itemize}
            \item \textbf{Viscosity.} We generate the data with $\mu$ scaled by $0.5$, $0.8$, $1.25$ and $2$, and reconstruct with the nominal $\mu$. The inlet velocity and $\alpha_{\max}$ are kept at their nominal values, so only the viscosity changes.
            \item \textbf{Diffusivity.} We generate the data with $D_{\max}$ scaled by the same factors, and reconstruct with the nominal $D_{\max}$. All other parameters, including $D_{\min}$, are unchanged.
            \item \textbf{Boundary location.} We keep the nominal data and instead displace the boundaries assumed by the reconstruction: either both inlets or the outlet are shifted by $\pm 2h$ along the boundary, where $h$ is the element size of the reconstruction mesh. The walls adjacent to the shifted boundary are moved accordingly.
        \end{itemize}
        The viscosity and diffusivity factors span a factor of two in each direction. They are chosen as a sensitivity range, not as a physiological range. All data are generated on the fine mesh and reconstructed on the coarse mesh, as in \cref{subsec:synthetic_data_generation}.
        
        Each reconstruction is scored against the truth that generated its data. The geometry metrics compare the recovered design with the true design. The velocity error $\varepsilon_u$ compares the recovered velocity with the velocity of the data-generating model, that is, at the true $\mu$ or $D_{\max}$. For the boundary-location tests, the reference is the nominal truth with the true boundary locations. To show how each mismatch changes the measurements, we also report the data change: the relative $L_2$ difference between the mismatched sinograms and the nominal sinograms. The results are summarized in \cref{tab:model_mismatch}.
        
        \begin{table}[htbp]
            \centering
            \caption{Reconstruction under model mismatch. The data change is the relative $L_2$ difference between the sinograms of the mismatched model and the nominal sinograms; it is zero for the boundary-location tests, whose data are nominal. $\mathbf{a}^\star$ lists the recovered scalings of the upper and lower inlets (true value $1$).}

            \label{tab:model_mismatch}
            \begin{tabular}{lccccc}
                \hline
                Case & Data change (\%) & DSC & $\mathrm{nRMSE}_\gamma$ & $\varepsilon_u$ (\%) & $\mathbf{a}^\star$ \\
                \hline
                Nominal (no mismatch) & 0 & 0.993 & 0.016 & 2.33 & (1.005, 0.996) \\
                \textbf{Viscosity $\mu$} & & & & & \\
                \hspace{1em}$\times 0.5$  & 4.67 & 0.982 & 0.036 & 3.99 & (0.995, 1.008) \\
                \hspace{1em}$\times 0.8$  & 1.57 & 0.994 & 0.017 & 2.11 & (1.000, 1.001) \\
                \hspace{1em}$\times 1.25$ & 1.63 & 0.990 & 0.020 & 3.03 & (1.004, 0.994) \\
                \hspace{1em}$\times 2$    & 5.27 & 0.987 & 0.032 & 4.69 & (1.002, 0.989) \\
                \textbf{Diffusivity $D_{\max}$} & & & & & \\
                \hspace{1em}$\times 0.5$  & 6.14 & 0.946 & 0.075 & 3.50 & (1.007, 0.987) \\
                \hspace{1em}$\times 0.8$  & 2.15 & 0.978 & 0.034 & 2.50 & (1.004, 0.994) \\
                \hspace{1em}$\times 1.25$ & 2.31 & 0.984 & 0.059 & 2.99 & (1.001, 1.000) \\
                \hspace{1em}$\times 2$    & 7.57 & 0.950 & 0.071 & 8.21 & (1.000, 0.994) \\
                \textbf{Inlet location} & & & & & \\
                \hspace{1em}$-2h$ & 0 & 0.991 & 0.032 & 9.55 & (0.992, 1.001) \\
                \hspace{1em}$+2h$ & 0 & 0.992 & 0.024 & 9.48 & (1.009, 0.989) \\
                \textbf{Outlet location} & & & & & \\
                \hspace{1em}$-2h$ & 0 & 0.991 & 0.018 & 2.98 & (1.007, 0.991) \\
                \hspace{1em}$+2h$ & 0 & 0.992 & 0.021 & 2.86 & (1.003, 0.994) \\
                \hline
            \end{tabular}
        \end{table}
        
        \textit{Viscosity.} The reconstruction is robust to a viscosity mismatch. Across the full factor-of-two range, the DSC stays at or above $0.982$ and $\varepsilon_u$ stays below $4.7\%$, compared with $0.993$ and $2.3\%$ without mismatch. The reason is that the data are only weakly sensitive to $\mu$: halving or doubling $\mu$ changes the sinograms by less than $5.3\%$. At the low Reynolds number of this case, with a prescribed inflow, the velocity field is set mainly by the lumen geometry and the inflow rate, and the contrast is carried by this velocity. The converse also holds: because the sinograms carry little information about $\mu$, the viscosity cannot be identified from them. A similar limitation is reported for velocity-image data in \cite{kontogiannis2022joint}. Quantities that depend directly on $\mu$, such as the pressure, therefore inherit the error of the assumed viscosity.
        
        \textit{Diffusivity.} A diffusivity mismatch is more impacting. At $\times 0.8$ and $\times 1.25$ the DSC remains above $0.978$, but at $\times 0.5$ and $\times 2$ it falls to $0.946$ and $0.950$, and $\mathrm{nRMSE}_\gamma$ rises to about $0.07$. These are also the cases with the largest data change ($6.1\%$ and $7.6\%$). Unlike the viscosity, the diffusivity directly controls how the contrast spreads, and hence the sinograms. With a wrong $D_{\max}$, one explanation is that the reconstruction can only match the observed spreading by altering the geometry. The diffusivity is therefore the transport parameter that most needs to be known, or estimated together with the geometry.
        
        \textit{Boundary location.} An outlet displaced by two elements has little effect: the DSC is $0.991$--$0.992$ and $\varepsilon_u$ is below $3\%$. The same displacement of the inlets leaves the geometry nearly unchanged (DSC $\geq 0.991$) but raises $\varepsilon_u$ to about $9.5\%$, because the prescribed inflow enters the domain at the wrong position.
        
        In all thirteen cases the recovered inlet scalings remain within $1.4\%$ of their true values. Overall, the geometry and the inflow are robust to errors in the viscosity and in the boundary locations of the order of the image resolution, while the contrast diffusivity must be known to within about $25\%$ for a DSC above $0.978$.
    
    \subsection{Sensitivity to numerical parameters}
    \label{subsection:numerical_parameters}
        
        \Cref{subsection:model_mismatch} varied the physical model. The reconstruction also depends on numerical parameters: the Brinkman penalty $\alpha_{\max}$, the solid diffusivity $D_{\min}$, the continuation of $q_\alpha$, $q_D$ and $\beta$, and the move limit of the optimizer. These values are chosen by the user, so a useful method should not depend strongly on them. We therefore change one parameter at a time in the reconstruction, keep the nominal data and all other settings of \cref{subsec:numerical_experiments_default}, and score each result against the nominal truth. The results are summarized in \cref{tab:numerical_parameters}. 
        
        \begin{table}[htbp]
            \centering
            \caption{Reconstruction under changes of the numerical parameters. Only the listed parameter differs from the nominal settings; the data are the nominal data in all cases. $\mathbf{a}^\star$ lists the recovered scalings of the upper and lower inlets (true value $1$).}
            \label{tab:numerical_parameters}
            \begin{tabular}{lcccc}
                \hline
                Case & DSC & $\mathrm{nRMSE}_\gamma$ & $\varepsilon_u$ (\%) & $\mathbf{a}^\star$ \\
                \hline 
                Nominal & 0.993 & 0.016 & 2.33 & (1.005, 0.996) \\
                \textbf{Brinkman penalty $\alpha_{\max}$} & & & & \\
                \hspace{1em}$\times 0.1$ & 0.852 & 0.168 & 10.23 & (0.971, 1.050) \\
                \hspace{1em}$\times 10$  & 0.988 & 0.036 & 8.47 & (1.004, 1.001) \\
                \textbf{Solid diffusivity $D_{\min}$} & & & & \\
                \hspace{1em}$\times 0.1$ & 0.994 & 0.015 & 2.41 & (1.004, 0.994) \\
                \hspace{1em}$\times 10$  & 0.991 & 0.019 & 2.72 & (1.000, 0.993) \\
                \hspace{1em}$\times 100$ & 0.966 & 0.052 & 5.34 & (0.991, 0.975) \\
                \textbf{Continuation} & & & & \\
                \hspace{1em}None (final values from iteration 1) & 0.991 & 0.017 & 2.35 & (0.999, 0.999) \\
                \hspace{1em}$q_\alpha$ ends at 25 & 0.988 & 0.024 & 7.38 & (1.004, 0.989) \\
                \hspace{1em}$q_\alpha = 100$ throughout & 0.974 & 0.044 & 5.76 & (0.998, 1.006) \\
                \hspace{1em}$q_D = 5$ throughout & 0.962 & 0.058 & 4.18 & (1.002, 0.990) \\
                \hspace{1em}$q_D$ ends at 20 & 0.966 & 0.055 & 3.25 & (1.005, 0.998) \\
                \hspace{1em}$\beta$ ends at 8 & 0.954 & 0.135 & 10.99 & (1.005, 0.998) \\
                \textbf{Move limit} & & & & \\
                \hspace{1em}$0.03$ & 0.992 & 0.017 & 2.28 & (1.004, 0.997) \\
                \hspace{1em}$0.08$ & 0.993 & 0.015 & 2.55 & (1.000, 0.996) \\
                \hline
            \end{tabular}
        \end{table}
        
        \textit{Brinkman penalty.} The penalty $\alpha_{\max}$ must be large enough that the solid blocks the flow, but not so large that the optimization becomes stiff. We set $\alpha_{\max} = 34{,}500\,\mathrm{g\,cm^{-3}\,s^{-1}}$, which is within $10\%$ of the value obtained from the rule given in Table~4 of \cite{alexandersen2023detailed}, $\alpha_{\max} = 10^4\alpha_{\mathrm{ref}}$ with $\alpha_{\mathrm{ref}} = 2.5\mu/w^2$ and $w$ the inlet width. A reconstruction run with that value, on data generated with it, gives the same result (DSC $0.993$).
        
        To measure how well the penalty approximates a no-slip wall, we compare the flow through the binary true design with a reference in which the solid elements carry a penalty of $10^9$. On the reconstruction mesh, the velocity differs by $1.15\%$ and the outlet flux by $0.04\%$, of the same order as the $1.2\%$ model error of the synthetic data (\cref{subsec:synthetic_data_generation}). Repeating the comparison on meshes of $40\times40$ and $120\times120$ elements gives velocity differences of $0.87\%$ and $1.32\%$, with the flow through the solid at about $10^{-3}$ of the inflow on all three meshes. The penalization error is therefore set by $\alpha_{\max}$ rather than by the element size; it grows slightly with refinement because the no-slip reference itself becomes more accurate.
        
        Reducing $\alpha_{\max}$ tenfold makes the walls permeable. The velocity differs from the no-slip reference by about $5.7\%$, and the reconstruction results in a lower DSC ($0.852$). Increasing it tenfold reduces that difference to $0.14\%$ and preserves the geometry (DSC $0.988$), but raises $\varepsilon_u$ to $8.5\%$. The nominal value therefore lies in the range where the walls are effectively impermeable and the geometry is recovered.
        
        \textit{Solid diffusivity.} A small positive $D_{\min}$ keeps the transport system well conditioned in the solid, at the cost of some artificial diffusion of contrast into the wall. We quantify this cost by its effect on the measurements. On the true design, changing $D_{\min}$ from its nominal value changes the sinograms by $0.09\%$ ($\times 0.1$), $0.83\%$ ($\times 10$) and $6.5\%$ ($\times 100$). The reconstruction follows the same pattern. Up to a factor of ten in either direction, the geometry is unchanged (DSC $0.991$--$0.994$) and $\varepsilon_u$ changes by at most $0.4$ percentage points. At $\times 100$, the artificial diffusion alters the data enough to degrade the reconstruction (DSC $0.966$).
        Because the artificial diffusion acts at the wall, we also repeated the stenosis reconstruction of \cref{subsection:stenotic_artery} with $D_{\min}$ increased tenfold. The narrowing is recovered equally well (DSC $0.997$) and the velocity error changes only from $2.55\%$ to $2.64\%$. A $D_{\min}$ within a factor of ten of the nominal value therefore does not impair the detection of a stenosis.
        
        \textit{Continuation.} The continuation gradually sharpens the interpolation and the projection over the first $100$ iterations. Removing it, so that the final values apply from the first iteration, changes the result little (DSC $0.991$). The schedule is therefore not what makes the reconstruction succeed; the final values are. A smaller final $q_\alpha$ ($25$ instead of $50$) or a fixed $q_\alpha = 100$ keeps the geometry (DSC $0.988$ and $0.974$) but increases $\varepsilon_u$ to $7.4\%$ and $5.8\%$. The final $q_D$ matters more: keeping it at $5$ or raising it to $20$, instead of $10$, lowers the DSC to $0.962$ and $0.966$. A sharper final projection ($\beta = 8$ instead of $4$) is unstable: a small region outside the true lumen stops changing at iteration $100$ and remains in the design, and the DSC drops to $0.954$. This is why the projection sharpness is capped at $\beta = 4$.
        
        \textit{Move limit.} The move limit of the optimizer has no measurable effect: limits of $0.03$, $0.05$ and $0.08$ give DSC $0.992$--$0.993$.
        
        In all cases the recovered inlet scalings remain within $5.0\%$ of their true values, and within $1.2\%$ in every case except $\alpha_{\max} \times 0.1$ and $D_{\min} \times 100$. Overall, the reconstruction is insensitive to the move limit, to $D_{\min}$ within a factor of ten, and to the use of continuation. It is sensitive to the final values of the interpolation parameters $q_\alpha$ and $q_D$, and it degrades when $\alpha_{\max}$ is too small or the projection too sharp.
        
    \subsection{Recovering Localized Luminal Narrowing in a Stenosed Artery}
    \label{subsection:stenotic_artery}

        The bifurcation tests whether the method recovers the branching of a vessel. A clinically relevant geometry is a local narrowing, a stenosis, as its severity is what guides treatment. Its severity affects the velocity field and the contrast transport \cite{maurya2025modelling, lu2024numerical}, and hemodynamic quantities such as pressure drop and fractional flow reserve \cite{nolte2022inverse, bovsnjak2025geometric}. This case tests whether the method can recover a stenotic feature from sparse projection data. 

\begin{figure}[htbp]
            \centering
                \includegraphics[scale=0.55,trim={0 0 0 0},clip]{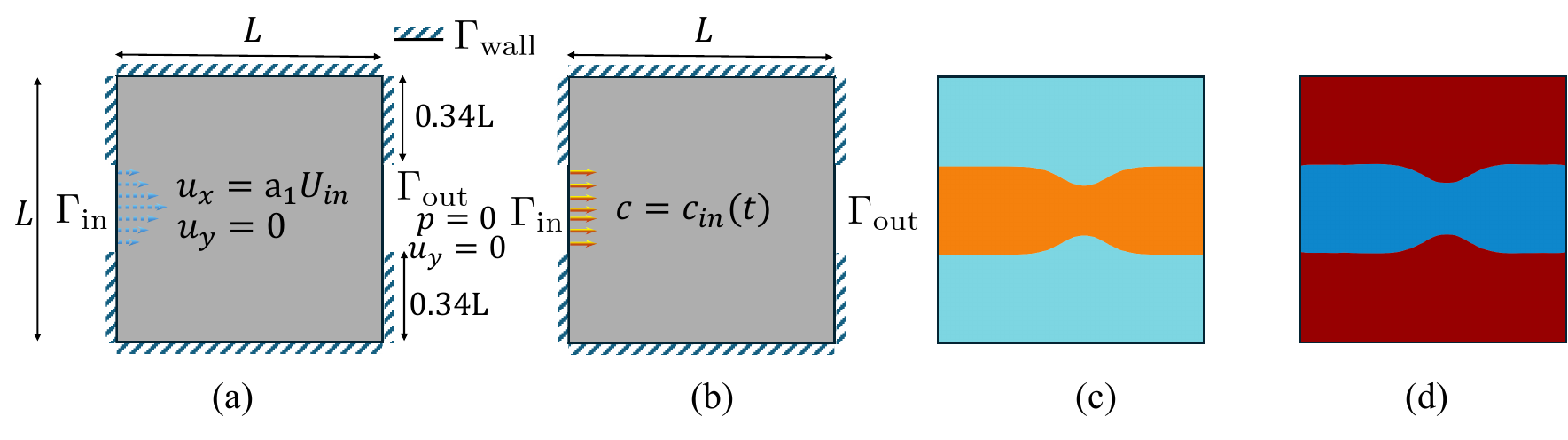}
                \caption{Arterial stenosis benchmark with 50\% lumen narrowing. 
                (a) Fluid flow boundary conditions.
                (b) Transient transport boundary conditions.
                (c) Ground-truth geometry used to generate the synthetic data. 
                (d) Recovered geometry obtained from the proposed framework using the sparse time-resolved sinogram data.}
                \label{fig:stenosis_geom_gt_op}
         \end{figure}

        The true geometry is a straight channel with a symmetric narrowing of Gaussian shape, similar to the profiles used in hemodynamics studies \cite{owasit2021mathematical}. It is a channel with a 50\% reduction in lumen height at the throat shown in \cref{fig:stenosis_geom_gt_op}(c). Blood enters through a single inlet on the left with a parabolic profile and leaves through an outlet on the right. Contrast enters through the inlet in two pulses, during $t \in [0.025, 0.525]\,\mathrm{s}$ and $[1.525, 2.025]\,\mathrm{s}$, each with smooth ramps of $0.05\,\mathrm{s}$, and the transport is simulated for $3\,\mathrm{s}$. All other settings, including the meshes, the time steps, the projection angles and the reconstruction parameters, are those of \cref{subsec:numerical_experiments_default,subsec:synthetic_data_generation}. The setup is shown in \cref{fig:stenosis_geom_gt_op}(a--b). Similar to the previous experiment (\cref{subsec:bifurcated_artery}), in this one we reconstruct both the pseudo-density field ($\boldsymbol{\gamma}$) and the inlet-flow scaling $a_1$.

        The reconstruction recovers the channel and the narrowing (\cref{fig:stenosis_geom_gt_op}(d)). The recovered geometry achieves $\mathrm{DSC} = 0.996$. The velocity error is $\varepsilon_u = 2.55\%$, and the recovered inlet scaling is $a^\star = 0.999$, within $0.2\%$ of its true value. The $\mathrm{nRMSE}_\gamma$ is  $0.093$.
        
\begin{figure}[htbp]
            \centering
                \includegraphics[scale=0.55,trim={0 0 0 0},clip]{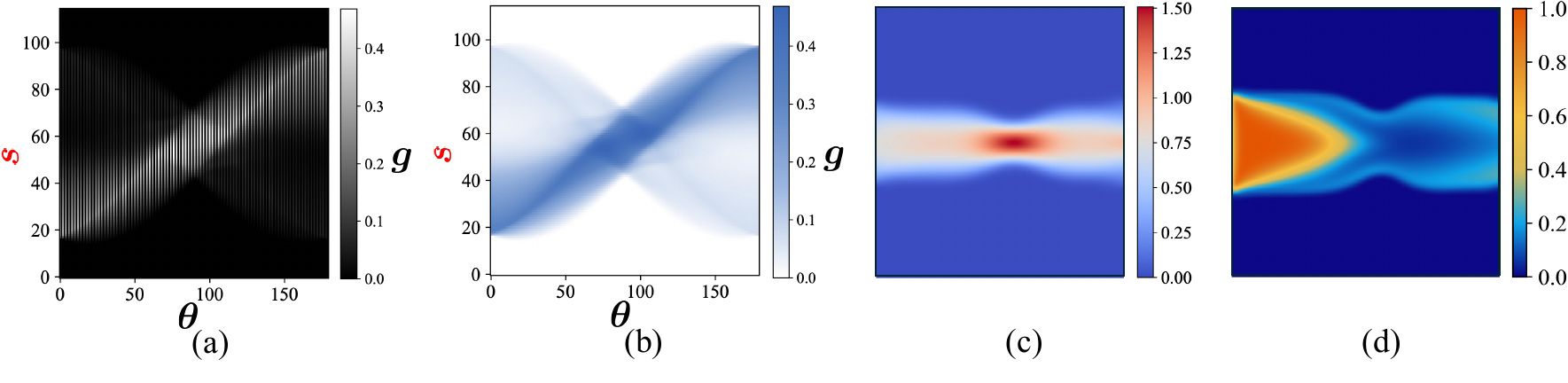}
                \caption{ Idealized symmetric arterial stenosis benchmark with 50\% lumen narrowing.
                (a) Sparse sinogram frame at $t=2.025\,\mathrm{s}$ generated from ground-truth geometry for synthetic data. 
                (b) Predicted sinogram frame generated from the concentration field at the same time. 
                (c) Predicted velocity-magnitude field. 
                (d) Predicted contrast concentration field at the same time. 
                }
                \label{fig:stenosis_forward}
         \end{figure}

       \Cref{fig:stenosis_forward} illustrates that the recovered solution explains the sparse sinogram data at $t=2.025\,\mathrm{s}$. The predicted sinogram in \cref{fig:stenosis_forward}(b), obtained by projecting the optimized concentration field, matches the sparse target in \cref{fig:stenosis_forward}(a). The associated velocity and concentration fields through the stenosis throat are shown in \cref{fig:stenosis_forward}(c,d).

    \subsection{Reconstruction of Branching Vascular Topology with Localized Stenosis}
    \label{subsection:carotid}
        The previous two cases test branching and narrowing separately. To evaluate the framework on a more challenging geometry, we combine both features in one geometry. The geometry is synthetic and resembles a carotid bifurcation. This bifurcation is associated with disturbed flow, low and oscillatory wall shear stress, and preferential plaque formation near the carotid sinus and outer wall regions \cite{zarins1983carotid, secomb2016hemodynamics, nolte2022inverse}. This makes the stenosed bifurcation a useful test case for evaluating the robustness of the proposed framework. 
        
\begin{figure}[htbp]
         	\centering
        		\includegraphics[scale=0.54,trim={0 0 0 0},clip]{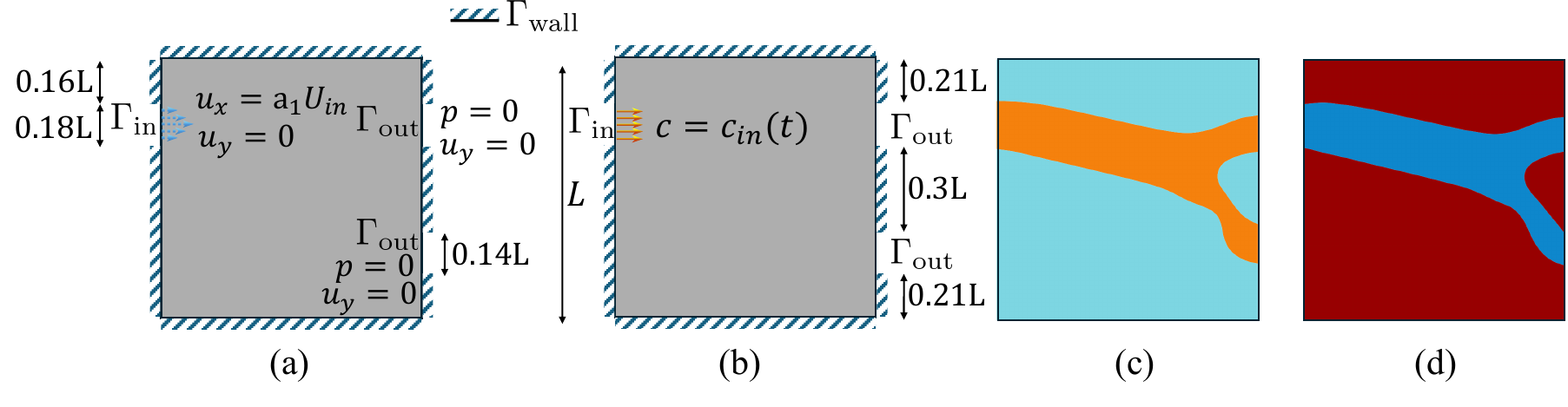}
         		\caption{Idealized carotid-bifurcation benchmark with an asymmetric stenosis in one daughter branch. 
                (a) Fluid flow boundary conditions.
                (b) Transient transport boundary conditions.
                (c) Ground-truth geometry used to generate the synthetic data. 
                (d) Recovered geometry obtained from the proposed physics-constrained reconstruction framework using sparse data.}
                \label{fig:carotid_geom_gt_op}
         \end{figure}

        This geometry has a single inlet on the left, that divides into an upper and a lower daughter branch, which leave through two outlets on the right. A plaque on the outer wall of the lower branch, just past the bifurcation, reduces the width of that branch. The contrast protocol is that of \cref{subsection:stenotic_artery}, and all other settings are those of \cref{subsec:numerical_experiments_default,subsec:synthetic_data_generation}. The setup is shown in \cref{fig:carotid_geom_gt_op}(a--b).

        From sparse projections, the reconstruction recovers the inlet scaling, both branches and the plaque (\cref{fig:carotid_geom_gt_op}(d)). The recovered geometry achieves DSC of $0.986$ and $\mathrm{nRMSE}_\gamma$ of $0.040$. The recovered inlet scaling is $0.999$, within $0.1\%$ of its true value, and the velocity error is $2.51\%$.

\begin{figure}[htbp]
            \centering
                \includegraphics[scale=0.55,trim={0 0 0 0},clip]{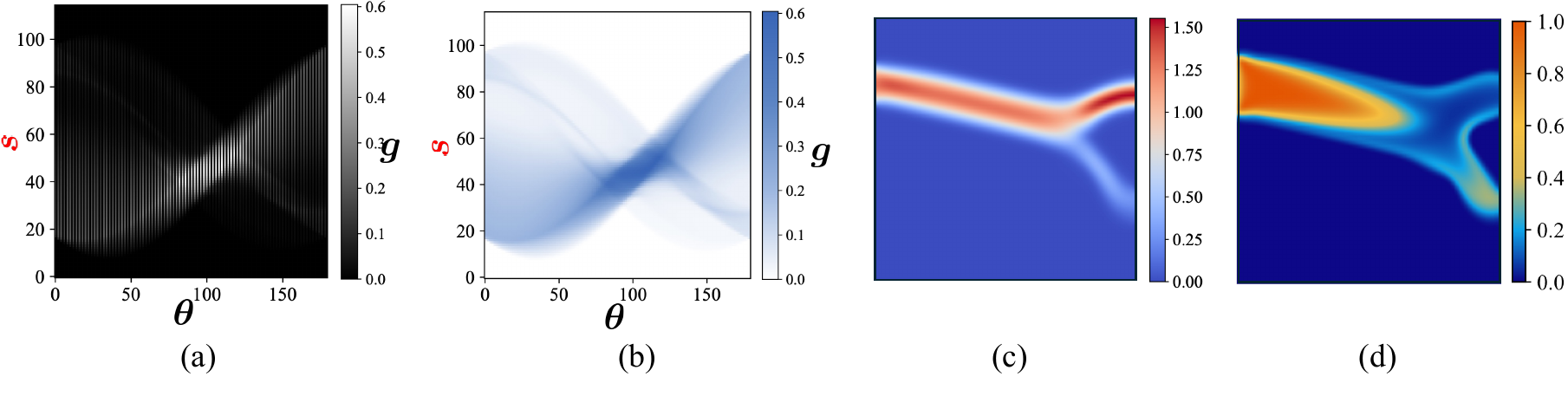}
                \caption{ Idealized carotid-bifurcation benchmark with an asymmetric stenosis in one daughter branch. 
                (a) Sparse-view sinogram frame at $t=2.025\,\mathrm{s}$ generated from ground-truth geometry for synthetic data. 
                (b) Predicted sinogram frame generated from the concentration field at the same time. 
                (c) Predicted velocity-magnitude field. 
                (d) Predicted contrast concentration field at the same time. 
                }
                \label{fig:carotid_forward_bc_all}
         \end{figure}

        \Cref{fig:carotid_forward_bc_all} shows that the recovered solution explains the sparse sinogram data: the predicted sinogram in \cref{fig:carotid_forward_bc_all}(b) matches the sparse target in (a), and (c,d) show the velocity and concentration fields through both daughter branches.

    \subsection{Robustness to Noisy Sparse-View Measurements}
    \label{sec:robustness_noise}

        The previous experiments used noise-free data at $90$ projection angles. Clinical CT data are corrupted by noise from the limited number of X-ray photons reaching the detector \cite{goldman2007principles, zhang2025artificial}, and this noise grows in low-dose or fast acquisitions, which also record fewer angles \cite{zhang2025artificial}. We therefore test the method with fewer angles (sparse) and noisy data, on the bifurcated artery of \cref{subsec:bifurcated_artery}.

        Following common practice in reconstruction studies, we add Gaussian noise to the synthetic sinograms \cite{guo2025computed, romanov2016simultaneous}. CT photon statistics are Poisson, but additive Gaussian noise is a common high-count approximation and is consistent with the squared $\mathcal{L}_2$ loss used here \cite{goldman2007principles, yoon2010simultaneous, romanov2016simultaneous, rautio2023learning}. Given the clean sinogram $g_{\mathrm{clean}}$, the noisy data are
        \begin{equation}
            g^{\mathrm{data}} = g_{\mathrm{clean}} + \delta \frac{\|g_{\mathrm{clean}}\|_2}{\|\boldsymbol{\xi}\|_2} \boldsymbol{\xi},
        \end{equation}
        
        where $\boldsymbol{\xi}$ has independent standard normal entries, one per frame, detector and angle, and $\delta$ is the relative noise level. The noisy sinogram is the fixed target of the objective; no noise is added to the predicted sinograms.
        
         We consider three settings of increasing difficulty:
        \begin{enumerate}
            \item \emph{sparse}: $18$ equally spaced angles (every tenth of $180$ projections), $\delta = 5\%$;
            \item \emph{ultra-sparse}: $10$ projections drawn at random from the $180$, $\delta = 5\%$;
            \item \emph{ultra-sparse, high noise}: as case 2, with $\delta = 25\%$.
        \end{enumerate}
        
        Because a single random draw could be favorable by chance, each case is repeated with three random seeds. In the ultra-sparse cases, a seed sets both the angle chosen and the noise; in the sparse case, the angles are fixed and the seed sets the noise. We report the mean and the standard deviation over the seeds. All other settings are those of \cref{subsec:numerical_experiments_default}, so the cost per iteration equals that of the noise-free case (\cref{subsection:mesh_dependency}).
        
        As an image-domain baseline, each frame of the same noisy sinograms is reconstructed by FBP \cite{van2014scikit}. The reconstructions are smoothed in time with a Gaussian filter, reduced to a temporal maximum-intensity projection \cite{smit2015timing}, and segmented with Otsu's threshold \cite{otsu1979athreshold}. The DSC of this segmentation is given in \cref{tab:noise_results}. 

        \begin{table}[htbp]
            \centering
            \caption{Reconstruction from sparse and noisy data: mean $\pm$ standard deviation over three seeds. Case~1 uses 18 equally spaced angles; cases~2 and~3 use 10 random angles drawn per seed. $\mathbf{a}^\star$ lists the recovered scalings of the upper and lower inlets (true value $1$). DSC (FBP) is the DSC of the filtered back-projection baseline on the same data.}
            \label{tab:noise_results}
            \begin{tabular}{lccccc}
                \hline
                Case & DSC & $\mathrm{nRMSE}_\gamma$ & $\varepsilon_u$ (\%) & $\mathbf{a}^\star$ & DSC (FBP)
                \\
                \hline
                90 angles, Noise-free  & 0.993 & 0.016 & 2.33 & (1.005, 0.996) & 0.959
                \\
                1: 18 angles, $5\%$ & $0.993 \pm 0.001$ & $0.051 \pm 0.001$ & $2.39 \pm 0.03$ & $(1.004, 0.995) \pm {<}0.001$ & $0.949 \pm {<}0.001$
                \\
                2: 10 angles, $5\%$ & $0.992 \pm 0.001$ & $0.047 \pm 0.008$ & $2.24 \pm 0.05$ & $(1.005, 0.997) \pm 0.001$ & $0.894 \pm 0.020$
                \\
                3: 10 angles, $25\%$ & $0.988 \pm 0.001$ & $0.088 \pm 0.008$ & $2.81 \pm 0.11$ & $(1.004, 0.996) \pm 0.003$ & $0.885 \pm 0.008$
                \\
                \hline
            \end{tabular}
        \end{table}
        
\begin{figure}
            \centering
                \includegraphics[scale=0.48,trim={0 0 0 0},clip]{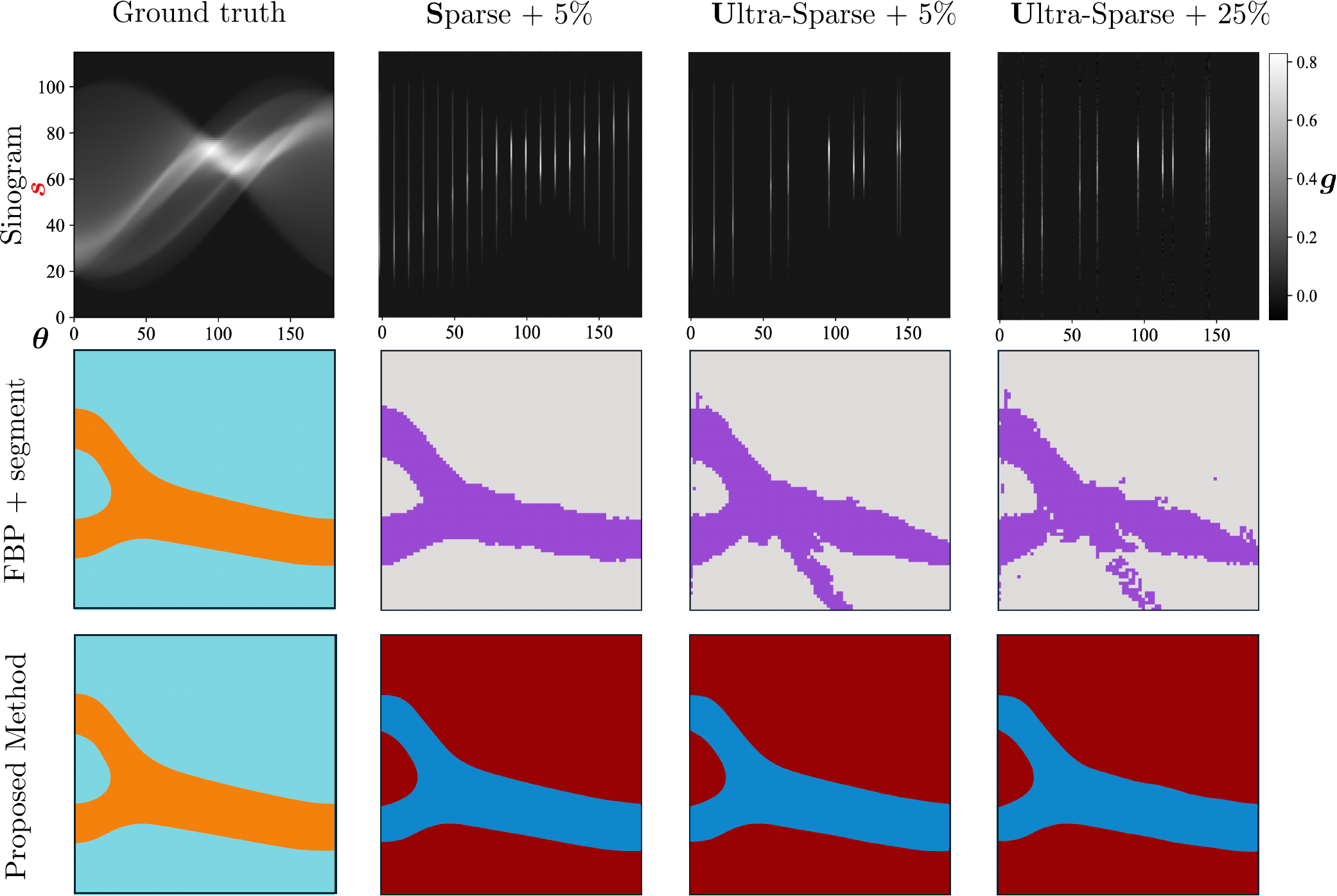}
                \caption{Reconstruction from noisy sparse-view sinogram measurements. The left column shows the ground-truth bifurcated geometry used to generate the synthetic data. The remaining columns compare three degraded measurement settings: uniform sparse data using 18 projections with $5\%$ relative Gaussian noise, ultra-sparse data using 10 randomly selected projections with $5\%$ relative Gaussian noise, and ultra-sparse data using 10 randomly selected projections with $25\%$ relative Gaussian noise. 
                Top row: representative noisy sinogram frames.
                Middle row: FBP, temporal maximum-intensity projection and Otsu segmentation.
                Bottom row: geometries recovered by the proposed physics-constrained topology optimization framework.
                }
                \label{fig:noisy_comparison}
        \end{figure}

        From the proposed method, the reconstruction degrades only slowly as the data worsen (\cref{tab:noise_results} and \cref{fig:noisy_comparison}). With $10$ angles and $25\%$ noise, the DSC falls from $0.993$ to $0.988$, the velocity error rises from $2.3\%$ to $2.8\%$, and the inlet scalings stay within $0.5\%$ of their true values on average. The spread over the seeds is small: the standard deviation of the DSC is at most $0.0013$. The result therefore does not depend on a favorable choice of angles or noise. 

        FBP degrades much faster. Even on the noise-free 90-angle data it reaches a DSC of only 0.959. It falls to 0.949 with 18 angles, to 0.894 with 10 angles and to 0.885 when the noise is raised to 25\%. With 10 angles, the FBP DSC also varies more between seeds ($\pm0.020$ against $\pm0.001$), and FBP recovers no velocity or inflow. 
        
        The reconstruction tolerates noise because it does not fit the noise. A reconstruction that fits the noise would reach a much lower objective than the true design, whose misfit is the noise itself. With $25\%$ noise, the objective at the recovered design is only $0.2\%$ lower than at the true design. Thus, the recovered design explains the signal, not the noise. This follows from the model: every candidate geometry must produce its sinograms through the flow and transport equations, which cannot reproduce random noise. The physics therefore acts as a regularizer, whereas an image-domain method such as FBP reconstructs each noisy frame directly. 

\section{Discussion}
\label{sec:discussion}
    
     The experiments show that the vessel geometry and the inflow can be recovered together from sinograms alone. Across the bifurcation, the stenosis and the carotid-like case, the recovered lumen matches the truth closely, the velocity error stays within a few percent, and the inlet scalings are recovered to within a fraction of a percent. Under sparse and noisy data the reconstruction degrades only slowly, while an image-domain baseline built from FBP degrades much faster (\cref{sec:robustness_noise}). The recovered design demonstrates that the physics prevents the reconstruction from fitting the noise. Rather than relying on classical image-domain segmentation, our approach enforces the governing flow and transport physics to regularize the inverse problem.
    
     The method assumes that the viscosity, the contrast diffusivity and the boundary locations are known. The reconstruction tolerates errors in the viscosity and in the boundary locations of the order of the image resolution, but not large errors in the diffusivity (\cref{subsection:model_mismatch}). The interpolation parameters $q_\alpha$ and $q_D$ also affect the result and were chosen empirically (\cref{subsection:numerical_parameters}). The optimizer is gradient-based and can be trapped in local minima: a sharper projection than the one used here freezes a spurious region early in the optimization (\cref{subsection:numerical_parameters}).
 
\section{Future work}
\label{sec:futurework}

    The present study is a two-dimensional proof of concept, and several of its modeling assumptions must be addressed before the framework can be applied clinically. The flow is steady, the blood is Newtonian, the vessel walls are rigid, and the projection is parallel-beam. The Newtonian assumption is common for arterial flow and blood is often treated as Newtonian in large vessels \cite{secomb2016hemodynamics, nolte2022inverse}. Non-Newtonian effects may be significant when the flow rate is low or when low shear rates occur locally \cite{secomb2016hemodynamics}. The shear rates in the present two-dimensional test cases are lower than typical arterial values, so the Newtonian model is a simplification here. The reconstruction is, however, insensitive to the viscosity level: halving or doubling it changes the DSC by at most $0.011$ (\cref{subsection:model_mismatch}). A uniform change is not the same as shear-thinning, but this insensitivity suggests that the recovered geometry depends weakly on the viscosity model. Non-Newtonian blood models have already been developed for density-based fluid topology optimization \cite{padhy2026toflux}, and will be included in the three-dimensional extension.

    The three-dimensional extension raises four challenges. First, the cost: the transport solve dominates the run time in the present 2D runs, and its size grows with the number of nodes and time steps. Second, the memory, which is dominated by the forward projection and the loss and grows with the number of elements, frames and angles. Third, mesh generation is simpler than in body-fitted CFD, because the design domain is a fixed box, but the box must resolve the smallest vessel of interest. Fourth, the boundary conditions: real vascular trees have several outlets with unknown flow division, and they exhibit secondary and helical flows that a two-dimensional model cannot represent. 
    
    Steady flow is a widely adopted baseline in PDE-constrained inverse problems and CT-based flow estimation \cite{sixou2020contrast, shusong2024deep}, and we adopt it here as well. It remains a simplification, since blood flow in coronary and carotid arteries is pulsatile. The pulsatile pressure also makes the arteries expand and contract, an effect that a rigid-wall model cannot represent. Future work will therefore extend the formulation to transient flow \cite{kreissl2011topology, deng2011topology}, include non-Newtonian blood models, for which density-based fluid topology optimization has already been developed \cite{padhy2026toflux}, and incorporate fluid--structure interaction to account for wall compliance \cite{nolte2022inverse}. The three-dimensional extension will also allow the reconstructed geometries to be validated against patient-specific in vivo data using SimVascular \cite{updegrove2017simvascular}.
        
    Besides these modeling assumptions, the framework assumes that several inputs are known. The viscosity, the contrast diffusivity and the inlet and outlet locations are prescribed rather than estimated. Future work will treat the contrast diffusivity and the boundary locations as unknowns and estimate them jointly with the vascular topology. The current framework also returns a single best-fit pseudo-density field and a single set of inlet scalings. A probabilistic extension would instead quantify the uncertainty in the recovered lumen, the boundary conditions, the transport properties and the derived hemodynamic quantities. Such an extension would show which vascular regions are well constrained by the sparse sinogram data and which remain ambiguous because of measurement noise, limited projection angles or uncertain model inputs. In structural TO, uncertain inputs have been treated with interval models in reliability-based and robust formulations \cite{li2025interval, li2025model}.

    These estimation problems rely on the optimizer, which is gradient-based and can converge to local minima. Future work will therefore investigate improved initialization and global-search strategies to reduce this sensitivity \cite{padhy2026tomatoes}. The parallel-beam projection operator is likewise a controlled idealization; scanner-specific acquisition geometries, including fan-beam CT, cone-beam CT and rotation-free architectures such as XMPI \cite{yao2025physics}, will be incorporated. Finally, estimating surface normals, wall shear stress, pressure drop and related near-wall biomarkers will require higher-resolution reconstructions, and efficient computation of these quantities will require surrogate models and high-performance computing.

\section{Conclusion}
\label{sec:conclusion}

     We have shown that vascular geometry and blood flow can be reconstructed directly from time-resolved contrast-enhanced CT projections, by formulating the reconstruction as a physics-constrained topology optimization problem. Instead of segmenting an image, the method searches for the geometry whose flow and contrast transport reproduce the measured sinograms. It recovers bifurcations and stenoses together with the inflow, from as few as ten projection angles and with up to $25\%$ noise, and it outperforms an image-domain baseline on the same data. The governing physics thus acts as a regularizer of the geometry. These results establish the method as a basis for vascular geometry and flow reconstruction from noisy and sparse CT data.

\section*{Funding}
The University of Wisconsin, Madison Graduate School supported this work. Roshan M. D'Souza acknowledges support from the National Science Foundation under Grant No. 2205265.

\section*{Author contributions}
RMD formulated the scientific problem. PT and RKP developed the computational framework. PT wrote the draft of the paper, with assistance from RKP. This was revised and edited by KS; additional inputs were provided by RMD.

\section*{Conflict of Interest}
The authors declare that they have no conflict of interest.

\section*{Ethics approval and Consent to participate}
Not applicable.

\section*{Data Availability and Replication of Results}
The Python code is available at \href{https://github.com/UW-ERSL/VASTO.git}{github.com/UW-ERSL/VASTO}.

\section*{Use of Generative AI}
During the preparation of this manuscript, the authors used Google's Gemini, OpenAI's ChatGPT and Anthropic's Claude to improve the language, readability and images. After using these tools, the authors reviewed and edited the content as needed and take full responsibility for the content of the publication.

\bibliographystyle{unsrt}  
\bibliography{references}  

\end{document}